\documentclass[12pt]{article}
\usepackage{amssymb,amsmath} 
\usepackage{graphicx}
\graphicspath{{graphics/}}
\usepackage{latexsym}
\usepackage{subfig}  
\usepackage{booktabs,setspace,caption}  
\usepackage{braket}
\usepackage{wasysym,verbatim}
\usepackage[numbers,sort&compress]{natbib}
\usepackage{textcomp}
\usepackage[utf8]{inputenc}

\usepackage{mathrsfs}
\usepackage[inner=2.9cm,outer=2cm]{geometry}
\usepackage[toc]{appendix}
\usepackage{color,soul} 
\usepackage{datetime}
\usepackage[
      colorlinks=true,
      linkcolor=darkblue,  
      urlcolor=darkblue,    
      filecolor=blue,     
      citecolor=myred, 
      linktocpage=true,
      pdfstartview=FitV,
      bookmarksopen=true    
      ]{hyperref}

\definecolor{darkblue}{rgb}{0, 0.3, 0.6}
\definecolor{myred}{rgb}{0.85,0,0}
\definecolor{darkgreen}{rgb}{0.2, 0.71, 0}

\newcommand{\eq}{\begin{equation}}
\newcommand{\qe}{\end{equation}}
\newcommand{\qm}{{\mathcal Q}_{-}}
\newcommand{\qp}{{\mathcal Q}_{+}}
\newcommand{\zm}{{\mathcal Z}_{-}}
\newcommand{\zp}{{\mathcal Z}_{+}}

\numberwithin{equation}{section}

\newenvironment{changemargin}[2]{%
\begin{list}{}{%
\setlength{\topsep}{0pt}%
\setlength{\leftmargin}{#1}%
\setlength{\rightmargin}{#2}%
\setlength{\listparindent}{\parindent}%
\setlength{\itemindent}{\parindent}%
\setlength{\parsep}{\parskip}%
}%
\item[]}{\end{list}}

\begin{document}  


\begin{titlepage}

\begin{flushright}
{\tt \small{IFT-UAM/CSIC-20-37}} \\
\end{flushright}

\vspace*{1.2cm}

\begin{center}
{\Large {\bf{Higher-derivative corrections to small black rings} }} \\

\vspace*{1.2cm}
\renewcommand{\thefootnote}{\alph{footnote}}
{\sl\large Alejandro Ruip\'erez}\footnotetext{E-mail: alejandro.ruiperezvicente@unipd.it}
\bigskip

 {Instituto de F\'{\i}sica Te\'orica UAM/CSIC\footnote{On leave.}\\
C/ Nicol\'as Cabrera, 13--15,  C.U.~Cantoblanco, E-28049 Madrid, Spain}\\ 

\vspace{0.5cm}

{Dipartimento di Fisica ed Astronomia ``Galileo Galilei", Università di Padova, \\
Via Marzolo 8, 35131 Padova, Italy}\\ 

\vspace{0.5cm}

{INFN, Sezione di Padova, \\
Via Marzolo 8, 35131 Padova, Italy}\\ 

\setcounter{footnote}{0}
\renewcommand{\thefootnote}{\arabic{footnote}}

\bigskip

\bigskip

\end{center}

\vspace*{0.1cm}

\begin{abstract}  
\begin{changemargin}{-0.95cm}{-0.95cm}
\noindent  
We compute the first-order $\alpha'$ corrections to a general family of supersymmetric heterotic backgrounds which describes a rotating superposition of a fundamental string and a momentum wave. We focus on a particular solution within this family that gives rise, upon dimensional reduction to five dimensions, to a black ring with two charges and one dipole that has a singular horizon with vanishing area in the supergravity approximation, better known as small black ring. We show that the singularity at the horizon persists even after the first-order $\alpha'$ corrections have been taken into account. This would imply that, contrarily to what has been argued in the literature, higher-derivative corrections do not resolve the singular horizon of the five-dimensional small black ring, at least at first order in $\alpha'$ expansion.
\end{changemargin}
\end{abstract} 

\end{titlepage}

\setcounter{tocdepth}{2}
{\small
\setlength\parskip{-0.5mm} 
\noindent\rule{15.7cm}{0.4pt}
\tableofcontents
\vspace{0.6cm}
\noindent\rule{15.7cm}{0.4pt}
}


\section{Introduction}

The study of black holes in string theory has been one of the major areas of research over the past decades, especially after Strominger and Vafa \cite{Strominger:1996sh} were able to reproduce the Bekenstein-Hawking entropy of a certain class of supersymmetric ---hence, extremal--- three-charge black holes in five dimensions by counting the degeneracy of BPS states of a system of D-branes wrapped on internal cycles. A crucial aspect of this setup is that the number of charges excited agrees with the minimum number of charges that are needed in order to have a supersymmetric black hole with non-vanishing horizon area, namely three in the case of five-dimensional black holes and four in the case of four-dimensional black holes.

There are, however, certain systems in string theory whose degeneracy of BPS states can be computed and it is finite but its low-energy (supergravity) description is an extremal black hole with vanishing horizon area.  In this case, it is widely accepted that the mismatch is solved by the higher-derivative corrections to the supergravity effective action, which must come to the rescue to stretch the area of the horizon in a way such that the corrected (Wald) entropy is in agreement with the degeneracy of BPS states. Perhaps the most representative example is the F1-P system, also known as Dabholkar-Harvey (DH) system, which consists of a heterotic fundamental string (F1) that is wound $w$ times along a compact direction, let us call it $z$, and a momentum wave (P) travelling also along $z$ with momentum $n$. This system contains a tower of half-BPS states whose degeneracy was computed in \cite{Dabholkar:1989jt} and the answer is

\begin{equation}
\label{eq:microentropyDH}
S_{\text{micro}}=4\pi\sqrt{nw}\, , \hspace{1cm} n, w>>1\, .
\end{equation}

\noindent
Supersymmetric black holes in four and five dimensions with the same conserved charges as the DH system were found in \cite{Sen:1994eb, Cvetic:1996xz} as particular cases of more general families of heterotic black holes. However, since only two charges are excited, the horizon area of these black holes vanishes in the supergravity approximation ---being this the reason why they are often called \emph{small black holes}--- and, what is even worse, the curvature blows up there. This actually informs us that the effective description breaks down near the horizon, where the curvature is no longer small in string units. It was then suggested by Sen \cite{Sen:1995in} that  higher-derivative corrections to the supergravity action\footnote{It turns out that in this case the relevant higher-derivative corrections are the so-called $\alpha'$-corrections. Loop or quantum corrections can be ignored at the horizon since the dilaton vanishes there.}, which are more and more relevant as we approach the singular horizon, could be able to stretch the horizon of the black hole. Furthermore, he was able to show  by means of a scaling argument that the correction to the entropy, if finite, would have the right dependence on the charges, i.e.:

\begin{equation}
S_{\rm{Wald}}\sim\sqrt{nw}\, .
\end{equation}

\noindent
The proportionality constant, however, could not be fixed at that time by just using the aforementioned scaling argument. Instead, its computation was addressed almost ten years later in \cite{Dabholkar:2004yr, Dabholkar:2004dq, Sen:2004dp, Hubeny:2004ji} for the case of four-dimensional heterotic small black holes, finding the precise of value of $4\pi$, in perfect agreement with (\ref{eq:microentropyDH}). However, it has been recently shown in \cite{Cano:2018hut} ---on the basis of an analytic solution--- that the regularization observed in \cite{Dabholkar:2004yr, Dabholkar:2004dq, Sen:2004dp, Hubeny:2004ji} is due to the presence of additional sources: solitonic 5-branes and Kaluza-Klein (KK) monopoles. Then, the regularized solution considered in those papers would not actually describe a genuine small black hole but just a special case of the regular black hole with four charges which has the property that its entropy coincides numerically with (\ref{eq:microentropyDH}).

Now, we would like to consider the case in which angular momentum is added to the F1-P system. In this case, the degeneracy of string states was computed by Russo and Susskind in \cite{Russo:1994ev} ---see also \cite{Bena:2004tk, Kraus:2005vz, Dabholkar:2005qs} and references therein--- and the result is

\begin{equation}
\label{eq:microentropy}
S_{\rm{micro}}=4\pi\sqrt{nw-JW}\,,
\end{equation}
where $J$ and $W$ are the angular momentum and the winding number along the direction of rotation.

Supergravity solutions with the same conserved charges as the rotating DH system were constructed in \cite{Callan:1995hn, Dabholkar:1995nc, Lunin:2001fv}.\footnote{They are reviewed  in Section~\ref{sec:smallblackrings}.} It was shown in \cite{Balasubramanian:2005qu, Dabholkar:2006za} that a particular class of them gives rise to supersymmetric two-charge black rings in five dimensions which also have a singular horizon with vanishing area, analogously to what occurs for small black holes. Then, the natural question was: do higher-derivative corrections stretch the horizon of small black rings? Some evidence in favor was given in \cite{Iizuka:2005uv}, where the entropy of five-dimensional small black rings was related to that of static four-dimensional small black holes by making use of the 4d-5d connection \cite{Gaiotto:2005gf, Elvang:2005sa, Gaiotto:2005xt, Bena:2005ni}. In addition to this, in \cite{Dabholkar:2006za} it was shown that the  scaling analysis used in \cite{Sen:1995in} for small black holes also applies for small black rings and that the Wald entropy would reproduce (\ref{eq:microentropy}) up to an overall proportionality constant.

\noindent
Let us observe, however, that the evidence in favor of the regularization of small black rings via higher-derivative corrections is based on certain premises that may not be necessarily true. On the one hand, the argument given in \cite{Dabholkar:2006za} is based on the regularization of four-dimensional small black holes, which has been refuted in \cite{Cano:2018hut}. On the other, the scaling argument of \cite{Sen:1995in, Dabholkar:2006za} only works if the correction to the Bekenstein-Hawking entropy is finite, which is not the case for small black holes ---see e.g.~\cite{Cano:2018qev, Faedo:2019xii}--- nor, presumably, for small black rings. 

Our contribution to the preceding discussion will be to compute analytically the leading higher-derivative corrections to the singular small black ring solution in five dimensions.

The rest of the paper is organized as follows: In Section~\ref{sec:effectiveaction}, we review the effective action of the heterotic superstring at first order in $\alpha'$, which captures the leading higher-derivative corrections to the small black ring solution. In Section~\ref{sec:genfamily}, we find an $\alpha'$-corrected family of solutions which describes, generically, a rotating superposition of a fundamental string and a momentum wave. In Section~\ref{sec:smallblackrings}, we discuss a particular class of solutions within this general family that leads, upon toroidal compactification, to the singular small black ring in five dimensions. Finally, Section~\ref{sec:conclusions} contains a summary of the results and a short discussion.


\section{The effective action of the heterotic superstring}
\label{sec:effectiveaction}

The purpose of this section is to review the relevant information about the effective action of the heterotic superstring that we shall use in the remaining of the text. We are going to assume that the string-loop (quantum) corrections can be safely ignored for our purposes.\footnote{This is expected to be the case for small black holes, for which the string coupling ($e^\phi$) vanishes at the horizon.}  Even in this limit, the effective action of the heterotic superstring contains an infinite tower of higher-derivative corrections to which we often refer as $\alpha'$ corrections, since a term with $2n$ derivatives will be multiplied by ${\alpha'}^{n-1}$, where $\alpha'=\ell^{2}_{s}$ and $\ell_{s}$ is the string length scale.\footnote{This is the unique dimensionful parameter of the theory.} The first terms in the $\alpha'$ expansion, up to  ${\cal O}(\alpha'^3)$, were determined in the late 1980s by different methods \cite{Gross:1986mw, Metsaev:1987zx, Bergshoeff:1989de}. For the purposes of this work, however, it is enough to incorporate only the leading curvature-squared (first-order in $\alpha'$) corrections. These were shown in \cite{Bergshoeff:1989de} to arise from the supersymmetrization of the Lorentz Chern-Simons term which has to be included in the (local) definition of the 3-form field strength $H$ (see Eq.~\eqref{eq:defH}) in order to cancel gauge and gravitational anomalies via the Green-Schwarz mechanism \cite{Green:1984sg}.\footnote{These are also the corrections which have been investigated in the past for heterotic small black holes (see for instance \cite{Dabholkar:2006za, Sen:2007qy,  Prester:2008iu, Prester:2010cw} and references therein), and therefore the ones which were argued to be responsible of the resolution of the horizon.} Using the conventions of \cite{Ortin:2015hya} and truncating the gauge fields, we have

\begin{equation}\label{eq:actionheterotic}
\begin{aligned}
S=\frac{g_s^2}{16\pi G_{\rm N}^{(10)}}\int d^{10}x \sqrt{-g}\,e^{-2\phi}\left \{R-4\left(\partial \phi\right)^2+\frac{1}{2\cdot 3!}H^2+\frac{\alpha'}{8}R_{(-)}{}_{\mu \nu ab}R_{(-)}{}^{\mu \nu ab}\right.\\
\left.+{\mathcal O}\left(\alpha'^2\right)\right\}\, ,
\end{aligned}
\end{equation}
where $G_{\rm N}^{(10)}$ is the ten-dimensional Newton constant and $g_{s}$ is the string coupling constant. The metric $g_{\mu\nu}$ is the string-frame metric, $\phi$ is the dilaton  and $H_{\mu\nu\rho}$ is the 3-form field strength of the Kalb-Ramond 2-form $B_{\mu\nu}$, whose definition is

\begin{equation}\label{eq:defH}
H=dB+\frac{\alpha'}{4}\omega^{\rm{L}}_{(-)}\, ,
\end{equation}
where $\omega^{\rm L}_{(-)}$ is the Lorentz Chern-Simons 3-form associated to the torsionful spin connection

\begin{equation}
\label{eq:cs3form}
\omega^{L}_{(-)}=d\Omega_{(-)}{}^{a}{}_{b}\wedge \Omega_{(-)}{}^{b}{}_{a}-\frac{2}{3} \Omega_{(-)}{}^{a}{}_{b}\wedge \Omega_{(-)}{}^{b}{}_{c}\wedge \Omega_{(-)}{}^{c}{}_{a}\, ,
\end{equation}
which in turn is defined as 

\begin{equation}
\Omega_{(-)}{}^{a}{}_{b}=\omega^{a}{}_{b}-\frac{1}{2}H_{c}{}^{a}{}_{b}\,e^c\, ,
\end{equation}
where $\omega^{a}{}_{b}$ represents the Levi-Civita spin connection. Finally, 

\begin{equation}
R_{(-)}{}^{a}{}_{b}=d\Omega_{(-)}{}^{a}{}_{b}-\Omega_{(-)}{}^{a}{}_{c}\wedge \Omega_{(-)}{}^{c}{}_{b}\, ,
\end{equation}
is the curvature 2-form associated to the torsionful spin connection.

\noindent
Let us notice two important aspects of the definition (\ref{eq:defH}). The first one is that it implies that the Bianchi identity of $H$ gets corrected by 

\begin{equation}
\label{eq:bianchi}
dH-\frac{\alpha'}{4}R_{(-)}{}^{a}{}_{b}\wedge R_{(-)}{}^{b}{}_{a}=0\, .
\end{equation}
The second one is that (\ref{eq:defH}) is a recursive definition that one has to implement order by order in $\alpha'$. Hence, the action (\ref{eq:actionheterotic})  and the Bianchi identity (\ref{eq:bianchi}) actually contain an infinite tower of implicit $\alpha'$ corrections. 

Finally, we want to emphasize that the action (\ref{eq:actionheterotic}) makes sense only in the limit where the higher-order $\alpha'$ corrections are subleading. This occurs, on general grounds, when the curvature scale of the solution $\cal R$ is small as compared to $\alpha'$, namely

\begin{equation}
\alpha' \mathcal {R}<<1\, .
\end{equation}

\noindent
If this is case, then it is justified to ignore terms with increasing number of derivatives since these will be more and more suppressed.


\subsection{Equations of motion}

In order to write the equations of motion derived from (\ref{eq:actionheterotic}), we shall use a lemma which was proven in \cite{Bergshoeff:1989de}. The lemma states that the variation of the action with respect to the torsionful spin connection $\delta S/\delta \Omega_{(-)}{}^a{}_{b}$ is proportional to $\alpha'$ and the zeroth-order equations of motion plus $\mathcal O\left(\alpha'^2\right)$ terms. Taking this into account, let us now separate the variation of the action with respect to the fields  into explicit and implicit variations (occurring through the torsionful spin connection) as follows 

\begin{equation}
\begin{aligned}
\delta S=&\frac{\delta S}{\delta e^{a}{}_{\mu}}\Bigg |_{\rm{exp}}\delta e^{a}{}_{\mu}+\frac{\delta S}{\delta \phi} \delta \phi+\frac{\delta S}{\delta B_{\mu\nu}}\Bigg |_{\rm{exp}}\delta B_{\mu\nu}\\
&+\frac{\delta S}{\delta \Omega_{(-)}{}^a{}_{b}}\left[\frac{\delta \Omega_{(-)}{}^a{}_{b} }{\delta e^{c}{}_{\rho}}\delta e^{c}{}_{\rho}+\frac{\delta \Omega_{(-)}{}^a{}_{b}}{\delta B_{\mu\nu}}\delta B_{\mu\nu}\right]\, .
\end{aligned}
\end{equation}
Because of the aforementioned lemma, if we work perturbatively in $\alpha'$, the second line above will yield ${\mathcal O}\left(\alpha'^2\right)$ terms which we shall ignore. Then, taking into account only the explicit variations, one has that the $\alpha'$-corrected equations of motion are reduced to the following set of equations

\begin{eqnarray}
\label{eq:einsteineom}
R_{\mu\nu}-2\nabla_{\mu}\partial_{\nu}\phi+\frac{1}{4}H_{\mu\rho\sigma}H_{\nu}{}^{\rho\sigma}&=&-\frac{\alpha'}{4}R^{(0)}_{(-)}{}_{\mu\rho ab}R^{(0)}_{(-)}{}_{\nu}{}^{\rho ab}+{\mathcal O}\left(\alpha'^2\right)\, , \\
\label{eq:dilatoneom}
\left(\partial \phi\right)^2-\frac{1}{2}\nabla^2\phi-\frac{1}{4\cdot 3!}H^2&=&\frac{\alpha'}{32}R^{(0)}_{(-)}{}_{\mu\nu ab}R^{(0)}_{(-)}{}^{\mu\nu ab}+{\mathcal O}\left(\alpha'^2\right)\, ,\\
\label{eq:KReom}
d\left(e^{-2\phi}\star H\right)&=&{\mathcal O}\left(\alpha'^2\right)\, ,
\end{eqnarray}

\noindent
where $R^{(0)}_{(-)}{}_{\mu\nu ab}$ denotes the curvature of the zeroth-order background. As we see, the corrected equations of motion are also of second-order in derivatives as the quadratic-curvature term in the action (\ref{eq:actionheterotic}) only acts as an effective source of energy and momentum.


\section{A family of $\alpha'$-corrected heterotic backgrounds}
\label{sec:genfamily}


\subsection{The zeroth-order solutions}

Let us consider the following field configuration at zeroth order in $\alpha'$

\begin{eqnarray}
ds^2&=&\frac{2}{{\mathcal Z}^{(0)}_{-}}du\left(dt+\omega^{(0)}-\frac{{\mathcal Z}^{(0)}_{+}}{2}du\right)-ds^2\left(\mathbb E^{d-1}\right)-ds^2\left(\mathbb T^{9-d}\right)\, ,\\
B&=&\frac{1}{{\mathcal Z}^{(0)}_{-}}\,du\wedge \left(dt+\omega^{(0)}\right)\, ,\\
e^{2\phi}&=&\frac{g_s^2}{{\mathcal Z}^{(0)}_{-}}\, ,
\end{eqnarray}
where  $ds^2\left({\mathbb T}^{9-d}\right)=dz^{1}dz^{1}+\dots +dz^{9-d}dz^{9-d}$ is the metric of a $(9-d)$-dimensional torus with total volume $V_{{\mathbb T}^{9-d}}=\left(2\pi \ell_{s}\right)^{9-d}$ and $ds^2\left({\mathbb E}^{d-1}\right)=dx^{m}dx^{m}$  is the metric of a $(d-1)$-dimensional Euclidean space on which the functions ${\mathcal Z}^{(0)}_{\pm}$ and the 1-form $\omega^{(0)}$ are defined. Hence, they do not depend neither on the light-cone coordinate $u=t-z$ nor on the internal coordinates parametrizing the torus.

The configuration we have just presented has been extensively studied in the literature\footnote{See for instance Refs.~\cite{Dabholkar:1990yf, Garfinkle:1992zj, Bergshoeff:1992cw, Bergshoeff:1994qm, Horowitz:1994rf, Callan:1995hn, Dabholkar:1995nc, Lunin:2001fv, Lunin:2001jy, Lunin:2002qf, Dabholkar:2006za}.}. It preserves half of the spacetime supersymmetries and generically describes, as we will see in the next section, a rotating superposition of a fundamental string and a momentum wave.

The zeroth-order equations of motion can be straightforwardly derived from Eqs.~(\ref{eq:einsteineom}), (\ref{eq:dilatoneom}) and (\ref{eq:KReom})  by just setting $\alpha'\rightarrow 0$. One finds that they are satisfied by our configuration if   

\begin{eqnarray}
\label{eq:0thorderZs}
\partial^2{\mathcal Z}^{(0)}_{\pm}&=&0\, ,\\
\label{eq:0thorderomega}
\partial_{p} \Omega^{(0)}{}_{pm}&=&0\, ,
\end{eqnarray}
where $\partial^{2}=\partial_{m}\partial_{m}$ and $\Omega^{(0)}=d\omega^{(0)}$.


\subsection{First-order $\alpha'$ corrections}

Let us assume that we have a solution to the zeroth-order equations of motion (i.e. two functions ${\mathcal Z}^{(0)}_{\pm}$ and a 1-form $\omega^{(0)}$ satisfying (\ref{eq:0thorderZs}) and (\ref{eq:0thorderomega}) respectively) and try to find a solution to the corrected equations of motion.  Then, we have to first write down an ansatz for the fields. The simplest possibility is to assume that the form of the solution will not be modified by the $\alpha'$ corrections. This is actually what happens in related systems \cite{Cano:2018qev, Chimento:2018kop, Cano:2018brq, Cano:2018hut}, where the  $\alpha'$ corrections only modify the components of the fields that are already active at zeroth order. We shall assume the same here. Then, our ansatz is 

\begin{eqnarray}
ds^2&=&\frac{2}{{\mathcal Z}_{-}}du\left(dt+\omega-\frac{{\mathcal Z}_{+}}{2}du\right)-ds^2\left(\mathbb E^{d-1}\right)-ds^2\left(\mathbb T^{9-d}\right)\, ,\\
B&=&\frac{1}{{\mathcal Z}_{-}}\,du\wedge \left(dt+\omega\right)\, ,\\
e^{2\phi}&=&\frac{g_s^2}{{\mathcal Z}_{-}}\, ,
\end{eqnarray}
with 

\begin{eqnarray}
{\mathcal Z}_{\pm}&=&{\mathcal Z}^{(0)}_{\pm}+\alpha'{\mathcal Z}^{(1)}_{\pm}+{\mathcal O}\left(\alpha'^2\right)\, ,\\
\omega&=&\omega^{(0)}+\alpha' \omega^{(1)}+{\mathcal O}\left(\alpha'^2\right)\, .
\end{eqnarray}

It is not difficult to see by using the results of Appendix~\ref{app:somecomputations} that the Lorentz Chern-Simons 3-form (\ref{eq:cs3form}) vanishes for our ansatz. Then, we find that the form of $H$ is  the same as in the zeroth-order case

\begin{equation}
H=dB=\frac{\partial_{m}\zm}{\zm^2}\, dx^{m}\wedge \left(dt+\omega\right)\wedge du-\frac{1}{\zm} \Omega\wedge du\, ,
\end{equation}
where $\Omega=d\omega$. Therefore, the left-hand side of equation of motion of $B_{\mu\nu}$ (\ref{eq:KReom}) has exactly the same form as for the zeroth-order solution. Since  the right-hand side vanishes, the conditions imposed by this equation are exactly those already found at zeroth order in $\alpha'$, namely

\begin{eqnarray}
\label{eq:firstorderzm}
\partial^{2}\zm&=&{\mathcal O}\left(\alpha'^2\right)\,, \\
\label{eq:firstorderomega}
\partial_{p}\Omega_{pm}&=&{\mathcal O}\left(\alpha'^2\right)\, .
\end{eqnarray}
It turns out that the equation of motion of the dilaton is also satisfied if (\ref{eq:firstorderzm}) holds. Additionally, for the Einstein equations (\ref{eq:einsteineom}) we find ---making use of the zehnbein basis defined in (\ref{eq:zehnbein}--- that the $+-$ and $+m$ components are satisfied if (\ref{eq:firstorderzm}) and (\ref{eq:firstorderomega}) hold whereas the $++$ component gives

\begin{equation}
\begin{aligned}
\frac{{\cal Z}^{(0)}_{-}}{2}\partial^2\zp=&-\alpha' R^{(0)}_{(-)}{}_{+mn+}R^{(0)}_{(-)}{}_{+mn-}+\frac{\alpha'}{4}R^{(0)}_{(-)}{}_{+mnp}R^{(0)}_{(-)}{}_{+mnp}+{\mathcal O}\left(\alpha'^2\right)\, \\
=&\alpha' {\cal Z}^{(0)}_{-} \left\{-\frac{1}{2}\left(\partial_{m}\partial_{n} {\cal Z}^{(0)}_{+}-\frac{\partial_{m}{\cal Z}^{(0)}_{+} \partial_{n}{\cal Z}^{(0)}_{-}}{{\cal Z}^{(0)}_{-}}\right)\left(\frac{\partial_{m}\partial_{n}{\cal Z}^{(0)}_{-}}{{\cal Z}^{(0)}_{-}}-\frac{\partial_{m} {\cal Z}^{(0)}_{-} \partial_{n}{\cal Z}^{(0)}_{-}}{({\cal Z}^{(0)}_{-})^2}\right)\right.\\
&\left.+\frac{1}{4}{\cal Z}^{(0)}_{-} \partial_{m}\left(\frac{\Omega^{(0)}_{np}}{{\cal Z}^{(0)}_{-}}\right)\partial_{m}\left(\frac{\Omega^{(0)}_{np}}{{\cal Z}^{(0)}_{-}}\right)\right\}+{\mathcal O}\left(\alpha'^2\right)\, .
\end{aligned}
\end{equation}
The above equation can be rewritten using that ${\cal Z}^{(0)}_{\pm}$ and $\omega^{(0)}$ satisfy the zeroth-order equations of motion as follows

\begin{equation}
\partial^2\left\{\zp-\alpha'\frac{\Omega^{(0)}{}_{mn}\Omega^{(0)}{}^{mn}-2\partial_{m}{\mathcal Z}^{(0)}_{+}\partial_{m}{\mathcal Z}^{(0)}_{-}}{4{\mathcal Z}^{(0)}_{-}}\right\}=\mathcal O\left(\alpha'^2\right)\, ,
\end{equation}
whose solution is

\begin{equation}
\label{eq:Zpcorrected}
\zp={\mathcal Z}^{(0)}_{+}+\alpha'\frac{\Omega^{(0)}{}_{mn}\Omega^{(0)}{}^{mn}-2\partial_{m}{\mathcal Z}^{(0)}_{+}\partial_{m}{\mathcal Z}^{(0)}_{-}}{4{\mathcal Z}^{(0)}_{-}}+\mathcal O\left(\alpha'^2\right)\, ,
\end{equation}
with ${\cal Z}^{(0)}_{+}$ harmonic in ${\mathbb E}^{d-1}$.  The remaining components turn out to be automatically satisfied for our ansatz. 

Therefore, we have just found that our ansatz is consistent with the equations of motion. A potential worry could be that the ansatz might not be general enough to provide an effective black-hole description of the DH states with angular momentum. This is studied in Appendix~\ref{app:susy}, where we show that turning on additional components of the fields breaks more supersymmetries than those preserved by the DH states.


\section{Small black rings from rotating strings}
\label{sec:smallblackrings}

Let us now discuss a particular class of heterotic backgrounds to which the results of Section~\ref{sec:genfamily} can be applied. These can be derived from the ones originally obtained in \cite{Callan:1995hn, Dabholkar:1995nc}, where also dependence in $u$ is allowed. The functions ${\mathcal Z}^{(0)}_{\pm}$ and the 1-form $\omega^{(0)}$ are given by

\begin{eqnarray}
{\mathcal Z}^{(0)}_{-}&=&1+\frac{q_-}{||x^m- F^m||^{d-3}}\, ,\\
{\mathcal Z}^{(0)}_{+}&=&1+\frac{q_{+}+q_{-}\dot F^2}{||x^m- F^m||^{d-3}}\, ,\\
\omega^{(0)}_{m}&=&\frac{q_- \dot F^{m}}{||x^m- F^m||^{d-3}}\, ,
\end{eqnarray}

\noindent
where $F^{m}=F^{m}(u)$ are arbitrary functions of $u=t-z$, $q_{-}$ and ${q}_{+}$ are constants and the dot denotes derivative with respect to $u$.

In the static limit, which corresponds to $F^{m}=\text{const}$, one recovers the solutions of \cite{Dabholkar:1990yf, Garfinkle:1992zj}, which describe a superposition of a fundamental string wrapped along the internal direction $z$ and a momentum wave that travels also along the $z$-direction. Upon dimensional reduction on ${\mathbb T}^{9-d}\times {\mathbb S}^{1}_{z}$, these static solutions correspond to extremal two-charge black holes in $d$ dimensions \cite{Cvetic:1996xz, Sen:1994eb}, which are also referred to as small black holes.

In the rotating case, $F^{m}\neq \text{const}$, the string is no longer located at a point in the non-compact space. Instead, its position is parametrically given by 

\begin{equation}
x^{m}=F^{m}(u)\, .
\end{equation}

\noindent
From this family of rotating string backgrounds, one can obtain a class of solutions with no dependence in the internal coordinate $z$ by the usual smearing procedure, which amounts to keeping only the zero mode in the Fourier expansion, namely

\begin{eqnarray}
{\mathcal Z}^{(0)}_{-}&=&1+\int_{0}^{\ell}\frac{q_{-}}{||x^m- F^m||^{d-3}}\, du\, ,\\
{\mathcal Z}^{(0)}_{+}&=&1+\int_{0}^{\ell} \frac{q_{+}+q_{-} \dot F^2}{||x^m- F^m||^{d-3}}\, du\, ,\\
\label{eq:omega0smeared}
\omega^{(0)}_{m}&=&\int_{0}^{\ell}\frac{q_-\dot F^{m} }{||x^m- F^m||^{d-3}}\, du\, ,
\end{eqnarray}

\noindent
where $\ell=2\pi w R_{z}$. 

\noindent
Let us explicitly check that the smearing procedure leads us to a solution. First, it is clear that ${\cal Z}^{(0)}_{\pm}$ are harmonic functions in ${\mathbb E}^{d-1}$ and therefore Eq.~(\ref{eq:0thorderZs}) is satisfied.\footnote{Except at the poles of the harmonic functions, where one must take into account the contributions from the sources \cite{Dabholkar:1995nc}.} It remains to check that Eq.~(\ref{eq:0thorderomega}) is also satisfied. Since $\omega^{(0)}_{m}$ is a harmonic function in ${\mathbb E}^{d-1}$, it reduces to 

\begin{equation}
\partial_{m}\partial_{p}\omega^{(0)}_{p}=0\, .
\end{equation}

\noindent
Using (\ref{eq:omega0smeared}), we have that 

\begin{equation}
\partial_{p}\omega^{(0)}_{p}=\frac{q_{-}}{||x^{m}-F^{m}\left(0\right)||^{d-3}}-\frac{q_{-} }{||x^{m}-F^{m}\left(\ell\right)||^{d-3}}\, ,
\end{equation}

\noindent
which vanishes if $F^{m}\left(u\right)=F^{m}\left(u+\ell\right)$, in which case (\ref{eq:0thorderomega}) is satisfied. We shall assume this in what follows. In fact, as in Refs.~\cite{Dabholkar:1995nc, Lunin:2001fv}, we are going to restrict ourselves to a circular profile of the form

\begin{equation}
F^{1}=R \cos \left( \frac{W u}{w R_{z}}\right)\, , \hspace{1cm} F^{2}=R\sin \left(\frac{W u}{w R_{z}}\right)\, , \hspace{1cm} F^{3}=\dots=F^{d-1}=0\, .
\end{equation}

\noindent
In this case, the string is wrapping a two-dimensional torus parametrized by the coordinates $z$ and $\psi$, the latter being the angular direction in the $x^{1}-x^{2}$ plane. The parameters $w$ and $W$ are telling us how many times the string is wound around the $z-$ and $\psi-$directions, respectively. As we are going to see next, a singular two-charge black ring in five dimensions is obtained when this solution is dimensionally reduced over ${\mathbb T}^{4}\times {\mathbb S}^{1}_{z}$.


\subsection{Five-dimensional small black rings}

The full five-dimensional configuration that one gets upon dimensional reduction on $\mathbb{T}^{4}\times \mathbb S^{1}_{z}$ is given in Appendix~\ref{app:dimensionalreduction}. Here we will just focus on the five-dimensional metric in the Einstein frame, which reads

\begin{equation}\label{eq:metric2chargebr}
ds^2_{\rm{E}, 5}=\left({\mathcal Z}^{(0)}_{+}{\mathcal Z}^{(0)}_{-}\right)^{-2/3}\left(dt+\omega^{(0)} \right)^2- \left({\mathcal Z}^{(0)}_{+}{\mathcal Z}^{(0)}_{-}\right)^{1/3}\,ds^2\left(\mathbb E^4\right)\, ,
\end{equation}

\noindent
where 

\begin{eqnarray}
\label{eq:zmbr}
{\mathcal Z}^{(0)}_{-}&=&1+\frac{\qm}{\ell}\int_{0}^{\ell}\frac{du}{||x^m- F^m||^{2}}\, ,\\
\label{eq:zpbr}
{\mathcal Z}^{(0)}_{+}&=&1+\frac{\qp}{\ell}\int_{0}^{\ell}\frac{du}{||x^m- F^m||^{2}}\, ,\\
\label{eq:omegabr}
\omega^{(0)}_{m}&=&\frac{\qm}{\ell}\int_{0}^{\ell}\frac{\dot F^{m}du}{||x^m- F^m||^{2}}\, ,
\end{eqnarray}

\noindent
having defined ${\cal Q}_{\pm}$ as

\begin{equation}
\qm=q_{-} \ell \, , \hspace{1cm}\qp={q}_+\ell+\frac{4\pi^2W^2R^2q_- }{\ell} \, .
\end{equation}

\noindent
As shown in \cite{Balasubramanian:2005qu, Dabholkar:2006za}, this  solution is nothing but a particular case of the supersymmetric three-charge black ring constructed in \cite{Elvang:2004ds, Bena:2004de, Gauntlett:2004qy}.  In order to see this explicitly, we have to perform the integrals appearing in Eqs.~(\ref{eq:zmbr}), (\ref{eq:zpbr}) and (\ref{eq:omegabr}), for which it is convenient to introduce the set of coordinates $\xi, \eta, \psi, \phi$, defined as follows

\begin{equation}
x^1=\xi \cos\psi\, , \quad x^2=\xi \sin\psi\, , \quad x^3=\eta \cos \phi\, ,\quad x^4=\eta \sin \phi\, .
\end{equation}

\noindent
After a bit of algebra, one finds 

\begin{eqnarray}
{\cal Z}^{(0)}_{\pm}&=&1+\frac{{\cal Q}_{\pm}}{\sqrt{\left(\xi^2+\eta^2+R^2\right)^2-4R^2\xi^2}}\, ,\\
\omega^{(0)}&=&\frac{\pi \qm W}{\ell}\,\left(\frac{\xi^2+\eta^2+R^2}{\sqrt{\left(\xi^2+\eta^2+R^2\right)^2-4R^2\xi^2}} -1\right)\, d\psi\, ,
\end{eqnarray}

\noindent
where we have used that

\begin{equation}
\int_{0}^{2\pi} \frac{\cos^n x\, dx}{1+a \cos x}=\frac{2\pi}{\sqrt{1-a^2}}\left(\frac{\sqrt{1-a^2}-1}{a}\right)^{n}\, .
\end{equation}

\noindent
We can write the solution in a more recognizable form by making use of the so-called ``ring coordinates'' $x$ and $y$, defined as \cite{Emparan:2001wn}

\begin{equation}
\xi=\frac{\sqrt{y^2-1}}{x-y}R\, , \quad \eta=\frac{\sqrt{1-x^2}}{x-y}R\, ,
\end{equation}

\noindent
where $-\infty\le y\le -1$ and $-1\le x\le1$. In terms of these coordinates, the four-dimensional Euclidean metric,  the functions ${\mathcal Z}^{(0)}_{\pm}$ and the 1-form $\omega^{(0)}$ are given by

\begin{eqnarray}
ds^2\left(\mathbb E^4\right)&=&\frac{R^2}{\left(x-y\right)^2}\left[\frac{dy^2}{y^2-1}+\left(y^2-1\right)d\psi^2+\frac{dx^2}{1-x^2}+\left(1-x^2\right)d\phi^2\right]\, ,\\
\label{eq:zssbr}
{\mathcal Z}^{(0)}_{\pm}&=&1+\frac{{\mathcal Q}_{\pm}}{2R^2}\left(x-y\right)\, ,\\
\label{eq:omegasbr}
\omega^{(0)}&=&-\frac{q}{2}\left(1+y\right)\,d\psi\, ,
\end{eqnarray}

\noindent
where we have defined

\begin{equation}
q=\frac{2\pi \qm W}{\ell}\, .
\end{equation}

\noindent
Once the solution is written in ring coordinates, it is easier to compare with the supersymmetric black ring of Refs.~\cite{Elvang:2004ds, Bena:2004de, Gauntlett:2004qy} and to confirm that it can be obtained by setting to zero one of the monopole charges and two of the dipole charges.


\subsubsection*{Physical parameters}

Let us discuss the physical interpretation of the four parameters $\qp, \qm, q$ and $R$ that, together with the asymptotic values of the scalars, determine the solution. To this aim, it is convenient to introduce a new pair of coordinates $\rho$ and $\theta$ such that

\begin{equation}
\label{eq:rhothetacoordinates}
\rho\sin\theta=R\frac{\sqrt{y^2-1}}{x-y}\, , \hspace{1cm} \rho \cos \theta=R\frac{\sqrt{1-x^2}}{x-y}\, .
\end{equation}

\noindent
In terms of these coordinates, the Euclidean metric, the functions ${\cal Z}^{(0)}_{\pm}$ and the 1-form $\omega^{(0)}$ read

\begin{eqnarray}
ds^2\left({\mathbb E}^4\right)&=&d\rho^2+\rho^2\left(d\theta^2+\sin^2\theta \,d\psi^2+\cos^2\theta \,d\phi^2\right)\, ,\\
\label{eq:Z0pm2}
{\mathcal Z}^{(0)}_{\pm}&=&1+\frac{{\mathcal Q}_{\pm}}{\Sigma}\, , \\
\label{eq:omega02}
\omega^{(0)}&=&-\frac{q}{2}\left(1-\frac{\rho^2+R^2}{\Sigma}\right)\, d\psi\, ,
\end{eqnarray}

\noindent
where we have defined

\begin{equation}
\Sigma=\sqrt{\left(\rho^2-R^2\right)^2+4R^2\rho^2\cos^2\theta}\, .
\end{equation}

The physical parameters can be identified by studying the large $\rho$-expansion of the fields. In particular, for the vector fields ---see Appendix~\ref{app:dimensionalreduction} for more details---, we have 

\begin{eqnarray}
A^{\pm}_{\underline{t}}&\sim& c_{\pm} \left[1-\frac{{\cal Q}_{\pm}}{\rho^2}+\mathcal O\left(\frac{1}{\rho^4}\right)\right]\, ,\\
A^{\pm}_{\underline{\psi}}&\sim& c_{\pm} \left[\frac{qR^2\sin^2\theta}{\rho^2}+\mathcal O\left(\frac{1}{\rho^4}\right)\right] \, ,\\
A^{0}_{\underline{\phi}}&\sim&c_{0}\left[\frac{qR^2\cos^2\theta}{\rho^2}+{\cal O}\left(\frac{1}{\rho^4}\right)\right]
\end{eqnarray}

\noindent
where $c_{0, +, -}$ are  moduli-dependent constants whose precise value is irrelevant for the present discussion. As we can see, the parameters $\qp$ and $\qm$ are the monopole electric charges whereas the parameter $q$, instead, controls the magnetic dipole charges. The relation between  these charges and the parameters that characterize the sources of the system ($n, w, W$) is \cite{Cano:2018qev, Dabholkar:2006za}

\begin{equation}
\qm=g_{s}^{2}\alpha' w\, , \hspace{1cm} \qp=\frac{g_{s}^{2}{\alpha'}^2}{R_{z}^2}n\, , \hspace{1cm} q=\frac{g_{s}^{2} \alpha' W}{R_{z}}\, .
\end{equation}

The mass and angular momenta of the solution can be identified by comparing the large $\rho$-expansion of the metric (\ref{eq:metric2chargebr})  with the Myers-Perry solution \cite{Myers:1986un}.\footnote{$\tilde \rho=\sqrt{\rho^2+\frac{\qp+\qm}{3}}$.} Doing so, we obtain

\begin{equation}
\begin{aligned}
ds^2_{\mathrm{E}, 5}\sim&\left(1-\frac{8 G_{\rm N}^{(5)} M}{3\pi \tilde \rho^2}\right)dt^2+ \frac{8 G_{\rm N}^{(5)}J_{\psi}\sin^2\theta}{\pi \tilde \rho^2} dt d\psi -  \left(1-\frac{8 G_{\rm N}^{(5)} M}{3\pi \tilde \rho^2}\right)d{\tilde \rho}^2\\
&-{\tilde \rho}^2\left(d\theta^2+\sin^2\theta d\psi^2+\cos^2\theta d\phi^2\right)\, ,
\end{aligned}
\end{equation}

\noindent
where 

\begin{eqnarray}
M&=&\,\frac{\pi \left(\qp+\qm\right)}{4 G_{\rm N}^{(5)}} ,\\
J_{\psi}&=&\frac{\pi q R^2}{4G_{\rm N}^{(5)}}\, .
\end{eqnarray}

\noindent
being $G_{\rm N}^{(5)}$ the five-dimensional Newton constant, $M$ the ADM mass and $J_{\psi}$ the angular momentum along the $\psi$-direction. We can now use the relation between the five- and ten-dimensional Newton constants and the expression of the latter in terms of the string moduli, $g_{s}$ and $\alpha'$,

\begin{equation}
\frac{1}{G_{\rm N}^{(5)}}=\frac{2\pi R_{z}\left(2\pi\ell_{s}\right)^4}{G_{\rm N}^{(10)}}\, , \hspace{1cm} G_{\rm N}^{(10)}=8\pi^6 g_{s}^{2}\alpha'^{4}\, ,
\end{equation}

\noindent
to rewrite the expressions for the mass and the angular momentum as follows

\begin{eqnarray}
M&=&\frac{n}{R_{z}}+\frac{R_{z}}{\alpha'}w\, ,\\
J_{\psi}&=&W \left(R/\ell_{s}\right)^2\, .
\end{eqnarray}
It is not difficult to see that the condition to avoid closed timelike curves reduces to 

\begin{equation}
J_{\psi}W<nw\, ,
\end{equation}
which provides a geometric derivation of the Regge bound \cite{Elvang:2004ds, Dabholkar:2005qs, Dabholkar:2006za}.


\subsubsection*{Singular horizon}

The supersymmetric three-charge black ring of Refs.~\cite{Elvang:2004ds, Bena:2004de, Gauntlett:2004qy} has a regular horizon at $y\rightarrow -\infty$ with topology ${\mathbb S}^{1}\times{\mathbb S}^{2}$. However, when only two monopole charges and one dipole are non-vanishing, the area of the horizon vanishes and, furthermore, one finds that the curvature diverges there, as it happens with static small black holes. Then, as we have anticipated, small black rings have a singular horizon with vanishing Bekenstein-Hawking entropy. Therefore, they cannot account for the microscopic degeneracy of the rotating Dabholkar-Harvey system (\ref{eq:microentropy}). Nevertheless, there is the possibility that the horizon is streched when higher-derivative corrections are taken into account, as proposed in \cite{Sen:1995in}. We shall study this possibility in the next section. 

\subsection{Higher-derivative corrections to small black rings}

Let us make use of the results of Section~\ref{sec:genfamily} to obtain the first-order $\alpha'$ corrections to the singular small black ring solution. The correction to the function $\zp$ is obtained by plugging Eqs.~(\ref{eq:zssbr}) and (\ref{eq:omegasbr}) into  Eq.~(\ref{eq:Zpcorrected}), which yields

\begin{equation}\label{eq:zpcorrectedsbr}
\begin{aligned}
\zp=&1+\frac{\qp}{2R^2}\left(x-y\right)+\frac{\alpha'}{4R^4} \frac{\left(x-y\right)^3\left[q^2R^2(x-y)+\qp \qm\left(x+y\right)\right]}{2R^2+\qm \left(x-y\right)}+\mathcal O\left(\alpha'^2\right)\\
=&1+\frac{\qp}{\Sigma}+2\alpha' \frac{q^2R^4-\qp \qm \rho^2}{\Sigma^3\left(\Sigma+\qm\right)}+\mathcal O\left(\alpha'^2\right),
\end{aligned}
\end{equation}
where in the second line we have made use of the coordinates $\rho$ and $\theta$ defined in Eq.~\eqref{eq:rhothetacoordinates}, which are more suitable to study, for instance, the asymptotic expansion of $\zp$, 

\begin{equation}
\begin{aligned}
\zp=&1+\frac{\qp}{\rho^2}+\frac{\qp R^2 \left(1-2 \cos^2\theta\right)}{\rho^4}\\
&+\frac{-2 \qp \qm \alpha'+ \qp R^4\left(1-6\cos^2\theta+6\cos^4\theta\right)}{\rho^6}+{\cal O}\left(\rho^{-8}\right)\, ,
\end{aligned}
\end{equation}
from where we can see that the higher-derivative corrections do not modify the momentum charge although they generically contribute to the higher-multipole moments.\footnote{As we show in Appendix~\ref{app:dimensionalreduction}, the Coulomb potential $\Phi^+=A^+_{\underline t}$  is related to the function $\zp$ by $\Phi^+=c_+\left(\zp^{-1}-1\right)$, where $c_+$ is a moduli-dependent constant, see Eq.~\eqref{eq:A+} for further details.}

As shown in Section~\ref{sec:genfamily}, the function $\zm$ and the 1-form $\omega$ remain uncorrected, so they are simply given by

\begin{eqnarray}
\label{eq:zmcorrectedsbr}
\zm&=&1+\frac{\qm}{2R^2}\left(x-y\right)+{\cal O}\left(\alpha'^2\right)\, ,\\
\label{eq:omegacorrectedsbr}
\omega&=&-\frac{q}{2}\left(1+y\right)\,d\psi+{\cal O}\left(\alpha'^2\right)\, .
\end{eqnarray}

The five-dimensional $\alpha'$-corrected metric  can be obtained from Eq.~(\ref{eq:metric2chargebr}) by simply replacing ${\cal Z}^{(0)}_{+}$, ${\cal Z}^{(0)}_{-}$ and $\omega^{(0)}$ by the expressions given in Eqs.~(\ref{eq:zpcorrectedsbr}), (\ref{eq:zmcorrectedsbr}) and (\ref{eq:omegacorrectedsbr}). As  occurs in the zeroth-order solution, the would-be horizon is at $y\rightarrow -\infty$, where the functions ${\cal Z}_{\pm}$ diverge. Expanding the $\alpha'$-corrected metric near $y\to-\infty$, we get 

\begin{equation}\label{eq:nearhorizon}
ds^2\sim \frac{f_H^2}{y^{8/3}}\left(dt^2- q y \,dtd\psi\right)-\frac{R^2}{f_H}\frac{dy^2}{y^{8/3}}-\frac{R^2}{f_H}\left(y^{-2/3}\,d\Omega^2_{(2)}+y^{4/3}\,d\psi^2\right)\, ,
\end{equation}
\noindent
 where 

\begin{equation}
f_{H}^{-3}=\frac{\alpha'}{\left(2R^2\right)^3}\left(q^2R^2-{\cal Q}_+{\cal Q}_-\right)\, .
\end{equation}

\begin{figure}[ht!]
\begin{center}
\includegraphics[scale=0.6]{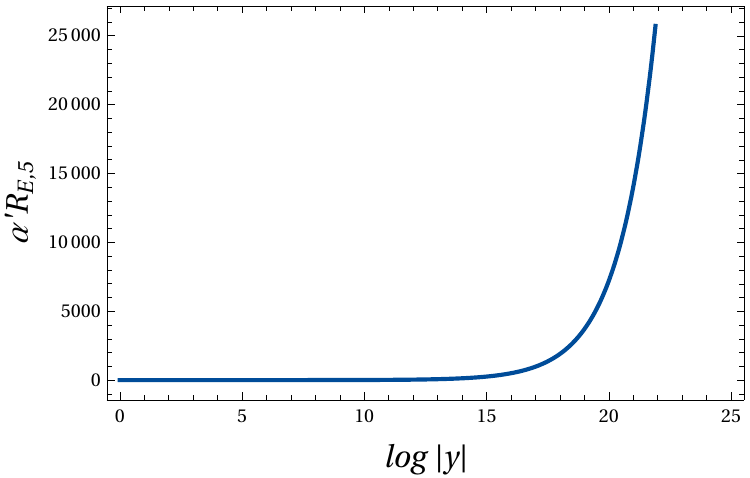}
\includegraphics[scale=0.58]{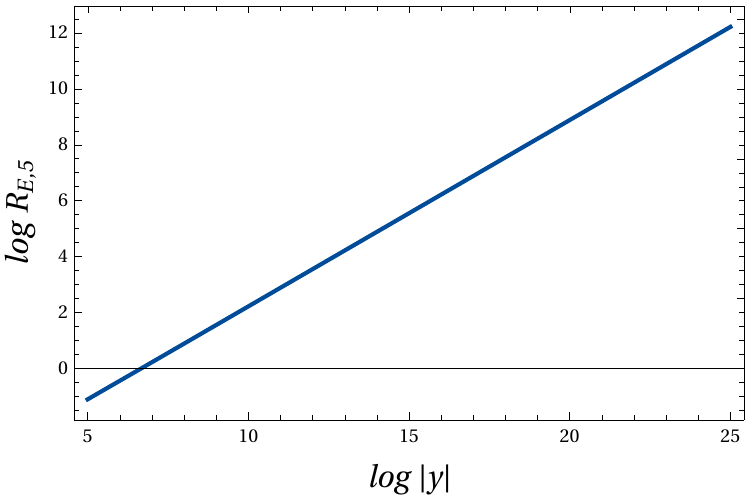}
\caption{In the plot on the left we have represented the five-dimensional Ricci scalar $R_{\rm{E},5}$ as a function of $\log |y|$ for $x=-1$ and for the particular values of the parameters: $R=10\sqrt{\alpha'}$, $\qp=\qm=2\cdot10^{3}\alpha'$ and $q=4\cdot10^{2}\sqrt{\alpha'}$. We find that the Ricci scalar diverges as $|y|^{2/3}$ as we approach $y\rightarrow-\infty$. This behaviour of the Ricci scalar near the singularity can be better appreciated in the plot on the right, where we have represented the logarithm of the Ricci scalar in the vertical axis.}
\label{fig:singularity}
\end{center}
\end{figure}

\noindent
It is clear that the near-horizon geometry is not ${\mathrm {AdS}}_{3}\times{\mathbb S}^2$ as in the supersymmetric three-charge black rings of \cite{Elvang:2004ds, Bena:2004de, Gauntlett:2004qy}. In fact, \eqref{eq:nearhorizon} describes a geometry with a curvature singularity at $y\to-\infty$.  In order to show this explicitly, we have represented  the Ricci scalar $R_{\rm{E}, 5}$ of the five-dimensional metric $g_{\rm{E}, 5}{}_{\mu\nu}$  as a function of $\log |y|$, see Fig.~\ref{fig:singularity}. Looking again at \eqref{eq:nearhorizon}, we see that the radius of the 2-sphere ${\mathbb S}^2$ vanishes in the $y\to -\infty$ limit as $R_{2}\sim |y|^{-1/3}$, while the radius of the circle ${\mathbb S}^{1}$ diverges as $R_{\psi}\sim y^{2/3}$. This implies that the product $R_{2}^{2}R_{\psi}$, which is proportional to the area of hypersurfaces of constant $t$ and $y$, is finite in the $y\to-\infty$ limit.\footnote{Moreover, the area is proportional to the quantity $\sqrt{q^2R^2-\qp \qm}$, which would be in agreement with the scaling argument of \cite{Sen:1995in, Dabholkar:2006za}. Let us note that the inequality $q^2R^2-{\cal Q}_+{\cal Q}_->0$ must hold in order to ensure that the metric has the right signature at the horizon. While this condition on its own is not enough to avoid closed timelike curves everywhere, we have checked numerically that there are values of the parameters for which these causal pathologies are absent.}

Before closing this section, let us study some interesting limits of the solution.

\subsubsection{$R\to0$ limit: static small black holes}

The five-dimensional small black holes studied in \cite{Cano:2018qev} can be obtained from our solution by first going to the coordinate system defined in Eq.~(\ref{eq:rhothetacoordinates}) and then taking the $R\rightarrow0$ limit. We get that the 1-form $\omega$  vanishes (so that the solution is static) and that $\zp$ and $\zm$ are given by

\begin{eqnarray}
\label{eq:zp1storder}
\zp&=&1+\frac{\qp}{\rho^2}-\frac{2\alpha' \qp \qm}{\rho^4\left(\rho^2+\qm\right)}+{\cal O}\left(\alpha'^2\right)\, ,\\
\label{eq:zm1storder}
\zm&=&1+\frac{\qm}{\rho^2}+{\cal O}\left(\alpha'^2\right)\, ,
\end{eqnarray}

\noindent
in agreement with the results of \cite{Cano:2018qev}. From a five-dimensional perspective, it seems that the metric develops a regular horizon with area 

\begin{equation}
A_{\rm{H}}=2\pi^2\sqrt{-2\alpha' \qp \qm}\, .
\end{equation}

\noindent
However, this solution cannot be trusted near the horizon since the curvature of the ten-dimensional metric diverges there and therefore the low-curvature assumption under which the action (\ref{eq:actionheterotic}) and the equations of motion derived from it were written is not satisfied. In five dimensions, this curvature singularity manifests in the divergence of the Kaluza-Klein scalar $k$ measuring the radius of the circle ${\mathbb S}^1_{z}$, 

\begin{equation}
k|_{\rho\to0}\sim \rho^{-3/2}\, ,
\end{equation}

\noindent
which informs us that the compactification is singular. 

\subsubsection{$R\to \infty$ limit}

Another interesting limit that we can study is that of infinite radius \cite{Elvang:2004ds, Elvang:2004rt}, which gives the $\alpha'$ corrections to the black string solutions of \cite{Bena:2004wv} when two of the dipoles and one of the charges are set to zero. In order to take this limit, we introduce new coordinates

\begin{equation}
r=-R/y\, , \hspace{1cm} \eta=R \psi\, ,
\end{equation}
and send $R\to \infty$ while keeping constant the charges ${\tilde {\cal Q}}_{\pm}$, defined as 

\begin{equation}
{\tilde {\cal Q}}_{\pm}\equiv \frac{{\cal Q}_{\pm}}{2R}\, .
\end{equation}
Doing so, we get  

\begin{eqnarray}
{\cal Z}_{+}&=&1+\frac{{\tilde {\cal Q}}_{+}}{r}+\frac{(q^2-4\tilde {\cal Q}_{+}\tilde {\cal Q}_{-})\alpha'}{8r^3(r+\tilde {\cal Q}_{-})}+{\cal O}(\alpha'^2)\, ,\\
{\cal Z}_{-}&=&1+\frac{\tilde {\cal Q}_{-}}{r}\, ,\\
\omega&=&\frac{q}{2r} d\eta \, .
\end{eqnarray}


\section{Discussion}
\label{sec:conclusions}

In this work, we have computed the leading higher-derivative corrections to a family of well-known supersymmetric heterotic backgrounds, focusing on a particularly interesting class of solutions within this general family which gives an effective (macroscopic) description of the DH states with angular momentum. In lower dimensions, it describes small black holes and black rings whose entropy is supposed to match the degeneracy of the DH states (\ref{eq:microentropy}) once the relevant higher-derivative corrections have been taken into account. Here we have shown that, contrary to these expectations, the first-order $\alpha'$ corrections dictated by the action \eqref{eq:actionheterotic} do not resolve the singular horizon of five-dimensional small black rings. Instead, we have found that the five-dimensional metric of the solution has a curvature singularity at the would-be horizon.

At this point, a reasonable question is if the higher-order corrections that we have neglected can play a relevant rôle in the regularization of small black holes. After all, the corrected solution that we have presented cannot be trusted near the singular horizon, where the perturbative treatment ceases to be justified and where presumably the higher-order corrections that we have ignored will not be subleading. Although this is certainly a possibility that deserves further study, we would not expect this to happen on general grounds. The reason to believe this is that one of the requirements that the solution must fulfill in order to be an effective description of the DH states is to preserve half of the supersymmetries, which  strongly constrains the form of the solution\footnote{See Appendix~\ref{app:susy}.} in a way which seems to be incompatible with having a regular horizon.

Finally, let us point out that our results, together with those of \cite{Cano:2018hut}, would indicate that the degeneracy of DH states cannot be accounted for solutions with a regular horizon. Alternatively, it might be possible to reproduce it by counting the degeneracy of fuzzball solutions, as proposed in  \cite{Mathur:2018tib}. This is a possibility that we would like to address in the future. Another task that is left for future work is to investigate the effect of higher-derivative corrections on the microstate geometries of the D1-D5 system \cite{Giusto:2004xm}, which are obtained from the solutions studied in Section~\ref{sec:smallblackrings} by performing a set of S- and T-duality transformations.


\section*{Acknowledgments}

I would like to thank Pablo A. Cano, Stefano Giusto, Tom\'as Ort\'in and Pedro F. Ram\'irez for valuable comments on an earlier version of this manuscript. This  work  has  been  supported  in  part  by a FPI-UAM predoctoral grant,  the  MCIU,  AEI,  FEDER (UE) grant PGC2018-095205-B-I00 and by the Spanish Research Agency (Agencia Estatal de Investigaci\'on) through the grant IFT Centro de Excelencia Severo Ochoa SEV-2016-0597.

\appendix


\section{Supersymmetry analysis}\label{app:susy}

The aim of this appendix is to show that the ansatz used in Section \ref{sec:smallblackrings} describes the most general half-supersymmetric superposition of fundamental strings and momentum waves with no dependence on the internal coordinate $z$ (where the string is wrapped) and, therefore, that it is general enough to provide an effective description of the DH states. The results of this appendix were first derived in \cite{Papadopoulos:2008rx}, which is based on earlier work of the author and collaborators \cite{Gran:2005wf, Gran:2007kh}. For the sake of convenience, here we will follow \cite{Fontanella:2019avn}, where it is presented an analysis the heterotic Killing spinor equations (KSEs) by using the spinor bilinear formalism.

\subsection{General form of the supersymmetric solutions}

According to Ref.~\cite{Fontanella:2019avn},  the metric of a supersymmetric configuration can always be written as

\begin{equation}
ds^2=2f\left(du+\beta\right)\left[dt+K \left(du+\beta\right)+\omega\right]-h_{\underline{mn}}\, dx^{m}dx^{n}\, ,
\end{equation}

\noindent
where $\omega=\omega_{\underline m}dx^{m}$ and $\beta=\beta_{\underline m} dx^m$ are 1-forms on the eight-dimensional space spanned by the coordinates $x^{m}$ and $f$ and $K$ are functions defined on this space.\footnote{In general, the objects that occur in the metric can also depend on $u$. However, we are assuming no dependence on this coordinate in order to be able to perform a KK reduction over this internal direction, where fundamental strings are wrapped.}  Given this form for the metric, we find convenient to introduce the following zehnbein basis 

\begin{equation}
e^{+}=f\left(du+\beta\right)\, , \hspace{0.5cm} e^{-}=dt+K \left(du+\beta\right)+\omega\, , \hspace{0.5cm} h_{\underline {mn}}\,dx^{m}dx^{n}=e^{m}e^{n}\, \delta_{mn}\, ,
\end{equation}

\noindent
in which the components of the spin connection $\omega^a{}_b$ read\footnote{Our conventions are such that $de^{a}=+\omega^{a}{}_{b}\wedge e^b$, with $a, b=+,-, m$.}

\begin{eqnarray}
\label{eq:spinconnection1}
\omega_{+-}&=&-\frac{1}{2}\partial_{m}\log f \, e^{m}\, ,\\
\omega_{+m}&=&-f^{-1}\partial_m K \, e^{+}-\frac{1}{2}\partial_{m}\log f \,e^-+\frac{1}{2}\left(K d\beta+d\omega\right)_{nm}\, e^n\, ,\\
\omega_{-m}&=&-\frac{1}{2}\partial_{m}\log f \,e^++\frac{f}{2}\left(d\beta\right)_{nm}\, e^n\, ,\\
\omega_{mn}&=&\frac{1}{2}\left(K d\beta+d\omega\right)_{mn}e^{+}+\frac{f}{2}\left(d\beta\right)_{mn}\,e^{-}\,+\varpi_{pmn}\, e^{p} \,,
\end{eqnarray}

\noindent
where $\varpi_{mnp}$ is the spin connection associated to $h_{\underline {mn}}$, defined as $de^m=-\varpi_{mn}\wedge e^n$.

Supersymmetric configurations must fulfill several conditions which can be divided into two sets.  On the one hand, the torsionful spin connection $\Omega_{(+)}{}^a{}_{b}\equiv \omega^a{}_b+\frac{1}{2}H_{c}{}^a{}_b \, e^c$ must satisfy  

\begin{eqnarray}
\label{eq:Omega+1}
\Omega_{(+)}{}_{[ab]-}&=&0\, ,\\
\label{eq:Omega+2}
\Omega_{(+)}{}_{am-}&=&0\, ,\\
\label{eq:Omega+3}
\nabla_{(+)}{}_{a}\Omega_{mnpq}&=&0\, ,
\end{eqnarray}

\noindent
where $\Omega_{mnpq}$ is a 4-form which can be understood as a $\mathfrak{spin}(7)$ structure and satisfies the following properties\footnote{We have adopted the convention of \cite{Fontanella:2019avn} that indices with the same latin letter $m_{i}, n_{i}, \dots $ are fully antisymmetrized.}

  \begin{eqnarray}
   \label{eq:O1O1simple}
    \Omega_{m_{1}m_{2}m_{3}p}\Omega_{n_{1}n_{2}n_{3}p}
    & =&
      -9\Omega_{m_{1}m_{2}n_{1}n_{2}}\delta_{m_{3}n_{3}}
      +6\delta_{m_{1}m_{2}m_{3},\, n_{1}n_{2}n_{3}}\, ,
    \\
    \label{eq:O2O2noantisimplebis}
    \Omega_{m_{1}m_{2}p_{1}p_{2}} \Omega_{n_{1}n_{2}p_{1}p_{2}}
    & =&
      -4\Omega_{m_{1}m_{2}n_{1}n_{2}}
      +12\delta_{m_{1}m_{2},\, n_{1} n_{2}}\, ,
  \\
      \label{eq:O2O2antisimplebis}
    \Omega_{m_{1}n_{2}p_{1}p_{2}} \Omega_{n_{1}m_{2}p_{1}p_{2}}
   & =&
      +4\Omega_{m_{1}m_{2}n_{1}n_{2}}
      +6\delta_{m_{1}m_{2},\, n_{1} n_{2}}\, .
    \\
    \label{eq:O2O2bis}
    \Omega_{m_{1}m_{2}n_{1}n_{2}} \Omega_{m_{3}m_{4}n_{1}n_{2}}
    & =&
      -4 \Omega_{m_{1}\cdots m_{4}}\, ,
    \\
    \label{eq:O3O3}
    \Omega_{mp_{1}p_{2}p_{3}}\Omega_{np_{1}p_{2}p_{3}}
    & =&
    42\delta_{mn}\, ,
    \\
      \label{eq:O2bis}
    \Omega_{m_{1}\cdots m_{4}}  \Omega_{m_{1}\cdots m_{4}}
    & \equiv&
      \Omega^{2}
       = 14\cdot 4!\, .
\end{eqnarray}

\noindent
In addition to Eqs.~\eqref{eq:Omega+1}, \eqref{eq:Omega+2} and \eqref{eq:Omega+3}, we have another set of conditions on the components of $H_{abc}$,

\begin{eqnarray}
\label{selfdualityH_{-mn}}
H^{(-)}_{-mn}&=&0\, ,\\
\label{selfdualityH_{+mn}}
H^{(-)}_{+mn}&=&\frac{1}{48}\Omega_{m}{}^{s_1 s_2s_3}\nabla_{+}\Omega_{n s_1s_2s_3}\, , \\
\label{selfdualityH_{mnp}}
H^{(-)}_{mnp}&=&\frac{1}{7}\left(2\partial_{q}\phi-H_{+-q}\right)\Omega^{q}{}_{mnp}\, ,
\end{eqnarray}

\noindent
where we have made use of the projectors acting on 2-forms $\Theta_{mn}$ and 3-forms $\Psi_{mnp}$ defined in \cite{Fontanella:2019avn}: 

\begin{eqnarray}
\Theta_{mn}&=&\Theta^{(+)}_{mn}+\Theta^{(-)}_{mn}\, , \hspace{1cm} \Theta^{(\pm)}_{mn}=\Pi^{(\pm)}_{mnpq}\Theta_{pq}\, , \\
\Psi_{mnp}&=&
\Psi^{(+)}_{mnp}+
\Psi^{(-)}_{mnp}\, , \hspace{1cm} 
\Psi^{(\pm)}_{mnp}=\Pi^{(\pm)}_{mnpqrs}
\Psi_{qrs}\, , 
\end{eqnarray}

\noindent
where

\begin{eqnarray}
\Pi^{(+)}_{mnpq}&=&\frac{3}{4}\left(\delta_{mn, pq}+\frac{1}{6}\Omega_{mnpq}\right)\, , \\
\Pi^{(-)}_{mnpq}&=&\frac{1}{4}\left(\delta_{mn, pq}-\frac{1}{2}\Omega_{mnpq}\right)\, , \\
\Pi^{(+)}_{m_1m_2m_3n_1n_2n_3}&=&\frac{6}{7}\left(\delta_{m_1m_2m_3, n_1n_2n_3}+\frac{1}{4}\Omega_{m_1m_2n_1n_2}\delta_{m_3 n_3}\right)\, , \\
\Pi^{(-)}_{m_1m_2m_3n_1n_2n_3}&=&\frac{1}{7}\left(\delta_{m_1m_2m_3, n_1n_2n_3}-\frac{3}{2}\Omega_{m_1m_2n_1n_2}\delta_{m_3 n_3}\right)\, .
\end{eqnarray}

Conditions \eqref{eq:Omega+1} and \eqref{eq:Omega+2} tell us that all the components of $\Omega_{(+)}{}_{ab-}$ vanish.\footnote{Although not constrained by these equations, $\Omega_{(+)}{}_{++-}=\omega_{++-}$ also vanishes, see \eqref{eq:spinconnection1}.} This implies that the components $H_{ab-}$ get fixed in terms of the objects that occur in the metric. We find, 

\begin{equation}\label{eq:H_{abc}}
H_{m+-}=\partial_{m}\log f\, , \hspace{1cm} H_{mn-}=f \left(d\beta\right)_{mn}\, .
\end{equation}

Before analyzing Eq.~\eqref{eq:Omega+3}, let us turn our attention to the second set of conditions. The first,  Eq.~\eqref{selfdualityH_{-mn}}, imposes 

\begin{equation}
\left(d\beta\right)^{(-)}_{mn}=0\, ,
\end{equation}

\noindent
so that the connection $\beta$ is that of an Abelian octonionic instanton \cite{Fubini:1985jm, Gunaydin:1995ku}.

The second one, Eq.~\eqref{selfdualityH_{+mn}}, can be rewritten by using \eqref{eq:O2O2noantisimplebis} and \eqref{eq:O3O3} as 

\begin{equation}
H^{(-)}_{+mn}=-2\, \omega^{(-)}_{+mn}=- (K d\beta+d\omega)^{(-)}_{mn}=-(d\omega)^{(-)}_{mn}\, .
\end{equation}

\noindent
Then, 

\begin{equation}
H_{+mn}=H^{(+)}_{+mn}-(d\omega)^{(-)}_{mn}=-(d\omega)_{mn}+ K (d\beta)_{mn}+f^{-1}\xi_{mn}\, ,
\end{equation}

\noindent
for some two-form $\xi=\frac{1}{2}\xi_{mn}\, e^m \wedge e^n$ such that $\xi^{(-)}_{mn}=0$. All in all, the general form of $H$ in supersymmetric configurations with no dependence on $z$ is

\begin{equation}
\begin{aligned}
H=&d\log f\wedge e^+ \wedge e^-+fe^-\wedge d\beta+e^{+}\wedge\left(-d\omega+K d\beta+f^{-1}\xi\right)\\
&+\frac{1}{3!} H_{mnp}\, e^{m}\wedge e^{n}\wedge e^{p}\, ,
\end{aligned}
\end{equation}

\noindent
with $H_{mnp}$ satisfying \eqref{selfdualityH_{mnp}}, which making use of \eqref{eq:H_{abc}} can be recasted as

\begin{equation}
H^{(-)}_{mnp}=\frac{1}{7}\partial_{q}\left(2\phi-\log f\right)\Omega^{q}{}_{mnp}\, .
\end{equation}

Now is the moment to analyze condition \eqref{eq:Omega+3}. Let us assume that there exists a basis $\{e^m\}$ in which the components of the 4-form $\Omega_{mnpq}$ are constant. In this basis, we have

\begin{equation}
\Omega_{(+)}{}_{a [m| s}\Omega_{s |npq]}=0\, ,
\end{equation}

\noindent 
and, hitting the above equation with $\Omega^{rnpq}$, we obtain the necessary and sufficient condition

\begin{equation}
\Pi^{(-)}_{mnpq}\Omega_{(+)}{}_{a pq}=0\, ,
\end{equation}

\noindent 
On the one hand, the $a=\pm$ components of the above equation are nothing but the selfduality conditions already derived for $d\beta$ and $\xi$. On the other, the $a=m$ component tells us that $\Omega_{(+)mnp}$ has special holonomy $G\subset\mathfrak{spin}(7)$.

Finally, note that the ansatz used in Section~\ref{sec:smallblackrings}
 is recovered for the particular choice

\begin{equation}
h_{\underline{mn}}=\delta_{\underline{mn}}\, , \hspace{1cm} H_{mnp}=0\, , \hspace{1cm}\xi_{mn}=0 \, , \hspace{1cm} \beta_{m}=0\, .
\end{equation}

\subsection{Killing spinor equations}

It was shown in \cite{Fontanella:2019avn} that the KSEs are satisfied by a constant spinor $\epsilon$ such that

\begin{eqnarray}
\label{firstconditionKS}
\Gamma^{+}\epsilon&=&0\, ,\\
\label{secondconditionKS}
\Pi^{(-)}\epsilon&=&0\, .
\end{eqnarray}

\noindent 
where 

\begin{equation}
\Pi^{(-)}=\frac{7}{8}\left(1-\frac{1}{336}\Omega_{m_{1}\dots m_{4}}\Gamma^{m_{1}\dots m_{4}}\right)\, .
\end{equation}

The first condition \eqref{firstconditionKS} kills half of the spacetime supersymmetries and, although in general half-supersymmetric configurations do not necessarily satisfy it, the class of half-supersymmetric solutions we are interested in (the one describing superpositions of fundamental strings with momentum) does, see for instance \cite{Papadopoulos:2008rx}. Then, if we want to preserve exactly this amount of supersymmetry, we must impose extra conditions on the fields to avoid imposing \eqref{secondconditionKS}, which would break additional supersymmetries. Let us study these conditions. To this aim, it is extremely useful to use the rewriting of the dilatino KSE provided in Eq. (3.46) of \cite{Fontanella:2019avn}: 

\begin{equation}
\left[\frac{1}{2}\left(2\partial_{m}\phi-H_{+-m}\right)\Gamma^{m}-\frac{1}{12}\left(H_{mnp}\Gamma^{mnp}+3H_{-mn}\Gamma^{-}\Gamma^{mn} \right)\right]\Pi^{(-)}\epsilon=0\, ,
\end{equation}

\noindent
where it has been already imposed \eqref{firstconditionKS}. Since we do not want to impose \eqref{secondconditionKS}, the term between brackets must vanish. Therefore,

\begin{equation}\label{conditionsdilatinoKSE}
\phi=\phi_{0}+\frac{1}{2}\log f\, , \hspace{1cm} H_{mnp}=0 \,, \hspace{1cm} \left(d\beta\right)_{mn}=0\, ,
\end{equation}

\noindent
where $\phi_{0}$ is an integration constant.

Let us now move to the gravitino KSE. For a constant spinor satisfying \eqref{firstconditionKS}, we only have to check that 

\begin{equation}
\Omega_{(+)}{}_{amn}\Gamma^{mn}\epsilon=0\, .
\end{equation}

As we have already seen, in a basis $\{e^{m}\}$ where the components of $\Omega_{m_{1}\dots m_{4}}$ are constant, we have that $\Omega^{(-)}_{(+)}{}_{amn}=0$. Therefore, we can use Eq. (A.46a) of  \cite{Fontanella:2019avn} to show that 

\begin{equation}\label{gravitinoKSE}
\Omega_{(+)}{}_{amn}\Gamma^{mn}\epsilon=\Omega_{(+)}{}_{amn}\Gamma^{mn}\Pi^{(-)}\epsilon=0\, ,
\end{equation}

\noindent
which implies that $\Omega_{(+)}{}_{amn}$ must vanish unless we impose \eqref{secondconditionKS}. This means

\begin{equation}
{\varpi}_{mnp}=0 \, , \hspace {1cm} \xi_{mn}=0\, ,
\end{equation}

\noindent
so that $e^m=dx^{m}$ and $h_{\underline{mn}}=\delta_{\underline{mn}}$. Therefore, we get\footnote{The 1-form $\beta$ can always be removed by the coordinate transformation $u\rightarrow u-\chi$, where $d\chi=\beta$.}

\begin{eqnarray}
ds^2&=&2e^{2\left(\phi-\phi_{0}\right)}du\left(dt+K du+\omega\right)-dx^{m}dx^{m}\, , \\
H&=&2e^{2\left(\phi-\phi_{0}\right)}d\phi\wedge du \wedge (dt+\omega)-f du\wedge d\omega\ ,
\end{eqnarray}

\noindent 
in agreement with \cite{Papadopoulos:2008rx}. This shows that the ansatz we used in Section~\ref{sec:smallblackrings} is the most general ansatz we can use to construct heterotic string backgrounds aimed to provide an effective description of the Dabholkar-Harvey states \cite{Dabholkar:1989jt, Russo:1994ev}.

\section{Connections and curvatures}
\label{app:somecomputations}

We work with the following zehnbein basis

\begin{equation}
\label{eq:zehnbein}
e^+=\mathcal Z_-^{-1} \, du\, , \quad e^-=dt-\frac{\mathcal Z_+}{2}du+ \omega\, , \quad e^{m}=dx^{m}\, , \quad e^{\alpha}=dz^{\alpha}\, ,
\end{equation}
where $m=1, \dots d-1$ and $\alpha=1, \dots, 9-d$. 

\subsection{Levi-Civita spin connection}

The non-vanishing components of the Levi-Civita spin connection, defined in our conventions as $de^a=\omega^a{}_b\wedge e^b$, are

\begin{eqnarray}
\omega_{+-}&=&\frac{1}{2}\partial_{m} \log \mathcal Z_- \, e^{m}\ ,\\
\omega_{+m}&=&\frac{1}{2}\partial_{m} \log \mathcal Z_- \, e^- -\frac{1}{2} \Omega_{mn}\, e^{n}+\frac{\mathcal Z_-}{2} \partial_{m} \mathcal Z_+\, e^+\\
\omega_{-m}&=&\frac{1}{2} \partial_{m} \log \mathcal Z_- \, e^+\\
\omega_{mn}&=&\frac{1}{2} \Omega_{mn}\, e^+\ .
\end{eqnarray}

\noindent
The curvature 2-form, defined as $R_{ab}=d\omega_{ab}-\omega_{ac}\wedge \omega^ {c}{}_{b}$, is given by

\begin{eqnarray}
R_{+-}&=& e^+\wedge e^- \left\{ \frac{\left(\partial\mathcal Z_-\right)^2}{4\zm^2}\right\}
+ e^{n}\wedge e^{+} \left\{-\frac{1}{4} \Omega_{np}\,\partial_p\log \mathcal Z_-\right\} \ , \\
R_{+m}&=&e^{n}\wedge e^{+}\left\{\frac{\zm}{2}\partial_{m}\partial_{n}\zp-\frac{1}{2}\partial_{(m|}\zp \partial_{|n)}\zm+\frac{1}{4}\Omega_{np}\Omega_{pm}\right\}
\nonumber\\
&&+e^{n}\wedge e^{p}\left\{\partial_{p}\Omega_{mn}+\Omega_{(n|p}\partial_{|m)}\log \zm\right\}+e^{+}\wedge e^{-}\left\{\frac{1}{4}\Omega_{mp}\frac{\partial_{p}\zm}{\zm}\right\}\\
\nonumber\\
&&+e^{n}\wedge e^{-}\left\{\frac{1}{2}\partial_{m}\partial_{n}\log \zm-\frac{1}{4}\frac{\partial_{m}\zm\partial_{n}\zm}{\zm^2}\right\}\, ,\\
R_{-m}&=& e^{n}\wedge e^{+}  \left\{\frac{1}{2}\frac{\partial_{m}\partial_{n}\zm}{\zm} -\frac{3}{4}\partial_{m} \log \zm \partial_{n}\log \zm\right\}\ , \\
R_{mn}&=&e^{p}\wedge e^{+}  \left\{\frac{\mathcal Z_-}{2}\partial_{p} \left(\zm^{-1} \Omega_{mn}\right)-\frac{1}{2}\Omega_{[m|p}\partial_{|n]}\log \zm\right\}\ .
\end{eqnarray}
\noindent
The Ricci tensor is 

\begin{eqnarray}
R_{++}&=& \frac{\zm}{2}\partial^{2}\zp-\frac{1}{2}\partial_{m}\zp \partial_{n}\zm-\frac{1}{4}\Omega^2\, ,\\
R_{+-}&=&\frac{1}{2}\frac{\partial^{2}\zm}{\zm} -\frac{\left(\partial \zm\right)^2}{\zm^2}\ ,\\
R_{+m}&=& \frac{\zm}{2}\partial_{n}\left(\zm^{-1}\Omega_{mn}\right)-\frac{1}{2}\Omega_{mp}\,\partial_{p}\log \zm\ ,\\
R_{mn}&=& -\frac{\partial_{m}\partial_{n}\zm}{\mathcal Z_-} +\frac{3}{2}\partial_{m}\log \zm \partial_{n}\zm\ .
\end{eqnarray}

Finally, the Ricci scalar is 

\begin{equation}
R=2R_{+-}-R_{mm}=2\frac{\partial^{2}\zm}{ \zm} -\frac{7}{2} \left(\partial\log \mathcal Z_-\right)^2\ .
\end{equation}


\subsection{Torsionful spin connection $\Omega_{(-)}{}_{ab}$}

The non-vanishing components of the torsionful spin connection are 

\begin{eqnarray}
\label{torsionfulspinconnection1}
\Omega_{(-)}{}_{+-}&=&\partial_{m} \log {{\mathcal Z}_-}\, e^{m}\ , \\
\label{torsionfulspinconnection2}
\Omega_{(-)}{}_{+m}&=&\frac{{\mathcal Z}_-}{2} \partial_{m} {\mathcal Z}_+\, e^+\ ,\\
\label{torsionfulspinconnection3}
\Omega_{(-)}{}_{-m}&=&\partial_{m}\log {{\mathcal Z}_-}\, e^+\ ,\\
\label{torsionfulspinconnection4}
\Omega_{(-)}{}_{mn}&=& \Omega_{mn}\, e^+\ .
\end{eqnarray}
\noindent
For the curvature 2-form, we get

\begin{eqnarray}
R_{(-)}{}_{+m}&=& e^n \wedge e^+ \,\, \left\{\frac{\zm}{2}\partial_{m}\partial_{n}\zp-\frac{1}{2}\partial_{m}\zp\partial_{n}\zm\right\}\ , \\
R_{(-)}{}_{-m}&=& e^n \wedge e^+ \,\, \left\{\frac{\partial_{m}\partial_{n}  {\mathcal Z}_{-}}{{\mathcal Z}_{-}}-\frac{\partial_{m}{\mathcal Z}_{-}}{{\mathcal Z}_{-}} \frac{\partial_{m}{\mathcal Z}_{-}}{{\mathcal Z}_{-}}\right\}\ , \\
R_{(-)}{}_{mn}&=& e^p\wedge e^+ \,\, \left\{\zm \partial_{p}\left(\frac{\Omega_{mn}}{\zm}\right)\right\}\ .
\end{eqnarray}

\section{Dimensional reduction to $d=5$ dimensions}
\label{app:dimensionalreduction}

It is well known that the dimensional reduction of ${\cal N}=1, d=10$ supergravity on ${\mathbb T}^4\times {\mathbb S}^{1}_{z}$ and truncation of all the Kaluza-Klein degrees of freedom of the 4-torus  ${\mathbb T}^{4}$ gives the STU model of ${\cal N}=1, d=5$ supergravity. The bosonic field content of the 5-dimensional theory is the metric, $g_{\rm{E},5}{}_{\mu\nu}$, two scalars, $\phi_{5}$ and $k$, and three vector fields, $A^{0,+,-}$. The explicit relation between the 10- and 5-dimensional fields in our conventions can be extracted from \cite{Cano:2017qrq}, and it is given by

\begin{eqnarray}
ds^2&=&e^{\phi-\phi_{\infty}}\left\{\left(k/k_{\infty}\right)^{-2/3}ds^2_{\mathrm{E,5}}-\left(k/k_{\infty}\right)^2\left(dz+\frac{k^{4/3}_{\infty}}{2\sqrt{3}}A^{+}\right)^2\right\}-ds^2\left(\mathbb{T}^{4}\right)\, , \\
H&=&-\frac{k^{2/3}_{\infty}}{\sqrt{3}g_{s}}e^{2\phi}k^{-4/3}\star_{5}F^{0}+\frac{k^{-2/3}_{\infty}}{\sqrt{3}g_s}F^{-}\wedge \left(dz+\frac{k^{4/3}_{\infty}}{2\sqrt{3}}A^{+}\right)\, , \\
\phi&=&\phi_{5}\, ,
\end{eqnarray}

\noindent
where $F^{0,+,-}=dA^{0,+,-}$ are the field strengths of the vector fields and $\star_{5}$ denotes the Hodge dual associated to the 5-dimensional metric. The moduli $\phi_{\infty}$ and $k_{\infty}$ are the asymptotic values of the dilaton $\phi$ and of the KK scalar $k$. They are related to the 10-dimensional moduli $g_{s}$ and $R_{z}$ by $g_{s}=e^{\phi_{\infty}}$ and $R_{z}=k_{\infty}\ell_{s}$.

For the configuration that we consider in the main text, the 5-dimensional fields are given by 

\begin{eqnarray}
ds^2_{\rm{E}, 5}&=&\left(\zp \zm\right)^2\left(dt+\omega\right)^2- \left(\zp \zm\right)^{1/3}ds^2\left(\mathbb{E}^{4}\right)\, ,\\
\label{eq:A+}
A^{+}&=& 2\sqrt{3}\, k_{\infty}^{-4/3}\,\frac{dt+\omega}{\zp}\, , \\
A^{-}&=&\frac{\sqrt{3}\,e^{\phi_{\infty}}\, k_{\infty}^{2/3}}{\zm}\left(dt+\omega\right)\, ,\\
A^{0}&=&-\sqrt{3}\,e^{-\phi_{\infty}}k_{\infty}^{2/3}\chi\, ,\\
e^{2\phi_{5}}&=&\frac{g^{2}_s}{\zm}\, , \hspace{1cm} k=k_{\infty} \,\frac{\zp^{1/2}}{\zm^{1/4}}\, ,
\end{eqnarray}

\noindent
where $\chi$ is a 1-form defined on ${\mathbb E}^4$ such that

\begin{equation}
d\chi=\star_{4}d\omega\, ,
\end{equation}

\noindent
where $\star_{4}$ is the Hodge dual with respect to the Euclidean metric. The integrability condition of $\chi$ is guaranteed to be satisfied because of the equation of motion of the Kalb-Ramond 2-form (\ref{eq:0thorderomega}).
\bibliographystyle{JHEP}
\bibliography{references}{}

\providecommand{\href}[2]{#2}\begingroup\raggedright\begin{thebibliography}{10}

\bibitem{Strominger:1996sh}
A.~Strominger and C.~Vafa, \emph{{Microscopic origin of the Bekenstein-Hawking
  entropy}}, \href{http://dx.doi.org/10.1016/0370-2693(96)00345-0}{\emph{Phys.
  Lett.} {\bf B379} (1996) 99--104},
  [\href{http://arxiv.org/abs/hep-th/9601029}{{\tt hep-th/9601029}}].

\bibitem{Dabholkar:1989jt}
A.~Dabholkar and J.~A. Harvey, \emph{{Nonrenormalization of the Superstring
  Tension}}, \href{http://dx.doi.org/10.1103/PhysRevLett.63.478}{\emph{Phys.
  Rev. Lett.} {\bf 63} (1989) 478}.

\bibitem{Sen:1994eb}
A.~Sen, \emph{{Black hole solutions in heterotic string theory on a torus}},
  \href{http://dx.doi.org/10.1016/0550-3213(95)00063-X}{\emph{Nucl. Phys.} {\bf
  B440} (1995) 421--440}, [\href{http://arxiv.org/abs/hep-th/9411187}{{\tt
  hep-th/9411187}}].

\bibitem{Cvetic:1996xz}
M.~Cvetic and D.~Youm, \emph{{General rotating five-dimensional black holes of
  toroidally compactified heterotic string}},
  \href{http://dx.doi.org/10.1016/0550-3213(96)00355-0}{\emph{Nucl. Phys.} {\bf
  B476} (1996) 118--132}, [\href{http://arxiv.org/abs/hep-th/9603100}{{\tt
  hep-th/9603100}}].

\bibitem{Sen:1995in}
A.~Sen, \emph{{Extremal black holes and elementary string states}},
  \href{http://dx.doi.org/10.1142/S0217732395002234}{\emph{Mod. Phys. Lett.}
  {\bf A10} (1995) 2081--2094},
  [\href{http://arxiv.org/abs/hep-th/9504147}{{\tt hep-th/9504147}}].

\bibitem{Dabholkar:2004yr}
A.~Dabholkar, \emph{{Exact counting of black hole microstates}},
  \href{http://dx.doi.org/10.1103/PhysRevLett.94.241301}{\emph{Phys. Rev.
  Lett.} {\bf 94} (2005) 241301},
  [\href{http://arxiv.org/abs/hep-th/0409148}{{\tt hep-th/0409148}}].

\bibitem{Dabholkar:2004dq}
A.~Dabholkar, R.~Kallosh and A.~Maloney, \emph{{A Stringy cloak for a classical
  singularity}},
  \href{http://dx.doi.org/10.1088/1126-6708/2004/12/059}{\emph{JHEP} {\bf 12}
  (2004) 059}, [\href{http://arxiv.org/abs/hep-th/0410076}{{\tt
  hep-th/0410076}}].

\bibitem{Sen:2004dp}
A.~Sen, \emph{{How does a fundamental string stretch its horizon?}},
  \href{http://dx.doi.org/10.1088/1126-6708/2005/05/059}{\emph{JHEP} {\bf 05}
  (2005) 059}, [\href{http://arxiv.org/abs/hep-th/0411255}{{\tt
  hep-th/0411255}}].

\bibitem{Hubeny:2004ji}
V.~Hubeny, A.~Maloney and M.~Rangamani, \emph{{String-corrected black holes}},
  \href{http://dx.doi.org/10.1088/1126-6708/2005/05/035}{\emph{JHEP} {\bf 05}
  (2005) 035}, [\href{http://arxiv.org/abs/hep-th/0411272}{{\tt
  hep-th/0411272}}].

\bibitem{Cano:2018hut}
P.~A. Cano, P.~F. Ram\'irez and A.~Ruip\'erez, \emph{{The small black hole
  illusion}},  \href{http://arxiv.org/abs/1808.10449}{{\tt 1808.10449}}.

\bibitem{Russo:1994ev}
J.~G. Russo and L.~Susskind, \emph{{Asymptotic level density in heterotic
  string theory and rotating black holes}},
  \href{http://dx.doi.org/10.1016/0550-3213(94)00532-J}{\emph{Nucl. Phys.} {\bf
  B437} (1995) 611--626}, [\href{http://arxiv.org/abs/hep-th/9405117}{{\tt
  hep-th/9405117}}].

\bibitem{Bena:2004tk}
I.~Bena and P.~Kraus, \emph{{Microscopic description of black rings in AdS /
  CFT}}, \href{http://dx.doi.org/10.1088/1126-6708/2004/12/070}{\emph{JHEP}
  {\bf 12} (2004) 070}, [\href{http://arxiv.org/abs/hep-th/0408186}{{\tt
  hep-th/0408186}}].

\bibitem{Kraus:2005vz}
P.~Kraus and F.~Larsen, \emph{{Microscopic black hole entropy in theories with
  higher derivatives}},
  \href{http://dx.doi.org/10.1088/1126-6708/2005/09/034}{\emph{JHEP} {\bf 09}
  (2005) 034}, [\href{http://arxiv.org/abs/hep-th/0506176}{{\tt
  hep-th/0506176}}].

\bibitem{Dabholkar:2005qs}
A.~Dabholkar, N.~Iizuka, A.~Iqubal and M.~Shigemori, \emph{{Precision
  microstate counting of small black rings}},
  \href{http://dx.doi.org/10.1103/PhysRevLett.96.071601}{\emph{Phys. Rev.
  Lett.} {\bf 96} (2006) 071601},
  [\href{http://arxiv.org/abs/hep-th/0511120}{{\tt hep-th/0511120}}].

\bibitem{Callan:1995hn}
C.~G. Callan, J.~M. Maldacena and A.~W. Peet, \emph{{Extremal black holes as
  fundamental strings}},
  \href{http://dx.doi.org/10.1016/0550-3213(96)00315-X}{\emph{Nucl. Phys.} {\bf
  B475} (1996) 645--678}, [\href{http://arxiv.org/abs/hep-th/9510134}{{\tt
  hep-th/9510134}}].

\bibitem{Dabholkar:1995nc}
A.~Dabholkar, J.~P. Gauntlett, J.~A. Harvey and D.~Waldram, \emph{{Strings as
  solitons and black holes as strings}},
  \href{http://dx.doi.org/10.1016/0550-3213(96)00266-0}{\emph{Nucl. Phys.} {\bf
  B474} (1996) 85--121}, [\href{http://arxiv.org/abs/hep-th/9511053}{{\tt
  hep-th/9511053}}].

\bibitem{Lunin:2001fv}
O.~Lunin and S.~D. Mathur, \emph{{Metric of the multiply wound rotating
  string}}, \href{http://dx.doi.org/10.1016/S0550-3213(01)00321-2}{\emph{Nucl.
  Phys.} {\bf B610} (2001) 49--76},
  [\href{http://arxiv.org/abs/hep-th/0105136}{{\tt hep-th/0105136}}].

\bibitem{Balasubramanian:2005qu}
V.~Balasubramanian, P.~Kraus and M.~Shigemori, \emph{{Massless black holes and
  black rings as effective geometries of the D1-D5 system}},
  \href{http://dx.doi.org/10.1088/0264-9381/22/22/010}{\emph{Class. Quant.
  Grav.} {\bf 22} (2005) 4803--4838},
  [\href{http://arxiv.org/abs/hep-th/0508110}{{\tt hep-th/0508110}}].

\bibitem{Dabholkar:2006za}
A.~Dabholkar, N.~Iizuka, A.~Iqubal, A.~Sen and M.~Shigemori, \emph{{Spinning
  strings as small black rings}},
  \href{http://dx.doi.org/10.1088/1126-6708/2007/04/017}{\emph{JHEP} {\bf 04}
  (2007) 017}, [\href{http://arxiv.org/abs/hep-th/0611166}{{\tt
  hep-th/0611166}}].

\bibitem{Iizuka:2005uv}
N.~Iizuka and M.~Shigemori, \emph{{A Note on D1-D5-J system and 5-D small black
  ring}}, \href{http://dx.doi.org/10.1088/1126-6708/2005/08/100}{\emph{JHEP}
  {\bf 08} (2005) 100}, [\href{http://arxiv.org/abs/hep-th/0506215}{{\tt
  hep-th/0506215}}].

\bibitem{Gaiotto:2005gf}
D.~Gaiotto, A.~Strominger and X.~Yin, \emph{{New connections between 4-D and
  5-D black holes}},
  \href{http://dx.doi.org/10.1088/1126-6708/2006/02/024}{\emph{JHEP} {\bf 02}
  (2006) 024}, [\href{http://arxiv.org/abs/hep-th/0503217}{{\tt
  hep-th/0503217}}].

\bibitem{Elvang:2005sa}
H.~Elvang, R.~Emparan, D.~Mateos and H.~S. Reall, \emph{{Supersymmetric 4-D
  rotating black holes from 5-D black rings}},
  \href{http://dx.doi.org/10.1088/1126-6708/2005/08/042}{\emph{JHEP} {\bf 08}
  (2005) 042}, [\href{http://arxiv.org/abs/hep-th/0504125}{{\tt
  hep-th/0504125}}].

\bibitem{Gaiotto:2005xt}
D.~Gaiotto, A.~Strominger and X.~Yin, \emph{{5D black rings and 4D black
  holes}}, \href{http://dx.doi.org/10.1088/1126-6708/2006/02/023}{\emph{JHEP}
  {\bf 02} (2006) 023}, [\href{http://arxiv.org/abs/hep-th/0504126}{{\tt
  hep-th/0504126}}].

\bibitem{Bena:2005ni}
I.~Bena, P.~Kraus and N.~P. Warner, \emph{{Black rings in Taub-NUT}},
  \href{http://dx.doi.org/10.1103/PhysRevD.72.084019}{\emph{Phys. Rev.} {\bf
  D72} (2005) 084019}, [\href{http://arxiv.org/abs/hep-th/0504142}{{\tt
  hep-th/0504142}}].

\bibitem{Cano:2018qev}
P.~A. Cano, P.~Meessen, T.~Ort\'in and P.~F. Ram\'irez,
  \emph{{$\alpha'$-corrected black holes in String Theory}},
  \href{http://dx.doi.org/10.1007/JHEP05(2018)110}{\emph{JHEP} {\bf 05} (2018)
  110}, [\href{http://arxiv.org/abs/1803.01919}{{\tt 1803.01919}}].

\bibitem{Faedo:2019xii}
F.~Faedo and P.~F. Ram\'irez, \emph{{Exact charges from heterotic black
  holes}}, \href{http://dx.doi.org/10.1007/JHEP10(2019)033}{\emph{JHEP} {\bf
  10} (2019) 033}, [\href{http://arxiv.org/abs/1906.12287}{{\tt 1906.12287}}].

\bibitem{Gross:1986mw}
D.~J. Gross and J.~H. Sloan, \emph{{The Quartic Effective Action for the
  Heterotic String}},
  \href{http://dx.doi.org/10.1016/0550-3213(87)90465-2}{\emph{Nucl. Phys.} {\bf
  B291} (1987) 41--89}.

\bibitem{Metsaev:1987zx}
R.~R. Metsaev and A.~A. Tseytlin, \emph{{Order alpha-prime (Two Loop)
  Equivalence of the String Equations of Motion and the Sigma Model Weyl
  Invariance Conditions: Dependence on the Dilaton and the Antisymmetric
  Tensor}}, \href{http://dx.doi.org/10.1016/0550-3213(87)90077-0}{\emph{Nucl.
  Phys. B} {\bf 293} (1987) 385--419}.

\bibitem{Bergshoeff:1989de}
E.~A. Bergshoeff and M.~de~Roo, \emph{{The Quartic Effective Action of the
  Heterotic String and Supersymmetry}},
  \href{http://dx.doi.org/10.1016/0550-3213(89)90336-2}{\emph{Nucl. Phys.} {\bf
  B328} (1989) 439--468}.

\bibitem{Green:1984sg}
M.~B. Green and J.~H. Schwarz, \emph{{Anomaly Cancellation in Supersymmetric
  D=10 Gauge Theory and Superstring Theory}},
  \href{http://dx.doi.org/10.1016/0370-2693(84)91565-X}{\emph{Phys. Lett. B}
  {\bf 149} (1984) 117--122}.

\bibitem{Sen:2007qy}
A.~Sen, \emph{{Black Hole Entropy Function, Attractors and Precision Counting
  of Microstates}},
  \href{http://dx.doi.org/10.1007/s10714-008-0626-4}{\emph{Gen. Rel. Grav.}
  {\bf 40} (2008) 2249--2431}, [\href{http://arxiv.org/abs/0708.1270}{{\tt
  0708.1270}}].

\bibitem{Prester:2008iu}
P.~Dominis~Prester and T.~Terzic, \emph{{$\alpha'$-exact entropies for BPS and
  non-BPS extremal dyonic black holes in heterotic string theory from
  ten-dimensional supersymmetry}},
  \href{http://dx.doi.org/10.1088/1126-6708/2008/12/088}{\emph{JHEP} {\bf 12}
  (2008) 088}, [\href{http://arxiv.org/abs/0809.4954}{{\tt 0809.4954}}].

\bibitem{Prester:2010cw}
P.~Dominis~Prester, \emph{{$\alpha'$-Corrections and Heterotic Black Holes}},
  1, 2010.
\newblock \href{http://arxiv.org/abs/1001.1452}{{\tt 1001.1452}}.

\bibitem{Ortin:2015hya}
T.~Ortin, \emph{{Gravity and Strings}}.
\newblock Cambridge Monographs on Mathematical Physics. Cambridge University
  Press, 2015,
  \href{http://dx.doi.org/10.1017/CBO9781139019750}{10.1017/CBO9781139019750}.

\bibitem{Dabholkar:1990yf}
A.~Dabholkar, G.~W. Gibbons, J.~A. Harvey and F.~Ruiz~Ruiz, \emph{{Superstrings
  and Solitons}},
  \href{http://dx.doi.org/10.1016/0550-3213(90)90157-9}{\emph{Nucl. Phys.} {\bf
  B340} (1990) 33--55}.

\bibitem{Garfinkle:1992zj}
D.~Garfinkle, \emph{{Black string traveling waves}},
  \href{http://dx.doi.org/10.1103/PhysRevD.46.4286}{\emph{Phys. Rev.} {\bf D46}
  (1992) 4286--4288}, [\href{http://arxiv.org/abs/gr-qc/9209002}{{\tt
  gr-qc/9209002}}].

\bibitem{Bergshoeff:1992cw}
E.~A. Bergshoeff, R.~Kallosh and T.~Ortin, \emph{{Supersymmetric string
  waves}}, \href{http://dx.doi.org/10.1103/PhysRevD.47.5444}{\emph{Phys. Rev.}
  {\bf D47} (1993) 5444--5452},
  [\href{http://arxiv.org/abs/hep-th/9212030}{{\tt hep-th/9212030}}].

\bibitem{Bergshoeff:1994qm}
E.~Bergshoeff, R.~Kallosh and T.~Ortin, \emph{{Black hole wave duality in
  string theory}},
  \href{http://dx.doi.org/10.1103/PhysRevD.50.5188}{\emph{Phys. Rev.} {\bf D50}
  (1994) 5188--5192}, [\href{http://arxiv.org/abs/hep-th/9406009}{{\tt
  hep-th/9406009}}].

\bibitem{Horowitz:1994rf}
G.~T. Horowitz and A.~A. Tseytlin, \emph{{A New class of exact solutions in
  string theory}},
  \href{http://dx.doi.org/10.1103/PhysRevD.51.2896}{\emph{Phys. Rev.} {\bf D51}
  (1995) 2896--2917}, [\href{http://arxiv.org/abs/hep-th/9409021}{{\tt
  hep-th/9409021}}].

\bibitem{Lunin:2001jy}
O.~Lunin and S.~D. Mathur, \emph{{AdS / CFT duality and the black hole
  information paradox}},
  \href{http://dx.doi.org/10.1016/S0550-3213(01)00620-4}{\emph{Nucl. Phys.}
  {\bf B623} (2002) 342--394}, [\href{http://arxiv.org/abs/hep-th/0109154}{{\tt
  hep-th/0109154}}].

\bibitem{Lunin:2002qf}
O.~Lunin and S.~D. Mathur, \emph{{Statistical interpretation of Bekenstein
  entropy for systems with a stretched horizon}},
  \href{http://dx.doi.org/10.1103/PhysRevLett.88.211303}{\emph{Phys. Rev.
  Lett.} {\bf 88} (2002) 211303},
  [\href{http://arxiv.org/abs/hep-th/0202072}{{\tt hep-th/0202072}}].

\bibitem{Chimento:2018kop}
S.~Chimento, P.~Meessen, T.~Ort\'in, P.~F. Ram\'irez and A.~Ruip\'erez,
  \emph{{On a family of $\alpha'$-corrected solutions of the Heterotic
  Superstring effective action}},
  \href{http://dx.doi.org/10.1007/JHEP07(2018)080}{\emph{JHEP} {\bf 07} (2018)
  080}, [\href{http://arxiv.org/abs/1803.04463}{{\tt 1803.04463}}].

\bibitem{Cano:2018brq}
P.~A. Cano, S.~Chimento, P.~Meessen, T.~Ort\'in, P.~F. Ram\'irez and
  A.~Ruip\'erez, \emph{{Beyond the near-horizon limit: Stringy corrections to
  Heterotic Black Holes}},
  \href{http://dx.doi.org/10.1007/JHEP02(2019)192}{\emph{JHEP} {\bf 02} (2019)
  192}, [\href{http://arxiv.org/abs/1808.03651}{{\tt 1808.03651}}].

\bibitem{Elvang:2004ds}
H.~Elvang, R.~Emparan, D.~Mateos and H.~S. Reall, \emph{{Supersymmetric black
  rings and three-charge supertubes}},
  \href{http://dx.doi.org/10.1103/PhysRevD.71.024033}{\emph{Phys. Rev.} {\bf
  D71} (2005) 024033}, [\href{http://arxiv.org/abs/hep-th/0408120}{{\tt
  hep-th/0408120}}].

\bibitem{Bena:2004de}
I.~Bena and N.~P. Warner, \emph{{One ring to rule them all ... and in the
  darkness bind them?}},
  \href{http://dx.doi.org/10.4310/ATMP.2005.v9.n5.a1}{\emph{Adv. Theor. Math.
  Phys.} {\bf 9} (2005) 667--701},
  [\href{http://arxiv.org/abs/hep-th/0408106}{{\tt hep-th/0408106}}].

\bibitem{Gauntlett:2004qy}
J.~P. Gauntlett and J.~B. Gutowski, \emph{{General concentric black rings}},
  \href{http://dx.doi.org/10.1103/PhysRevD.71.045002}{\emph{Phys. Rev. D} {\bf
  71} (2005) 045002}, [\href{http://arxiv.org/abs/hep-th/0408122}{{\tt
  hep-th/0408122}}].

\bibitem{Emparan:2001wn}
R.~Emparan and H.~S. Reall, \emph{{A Rotating black ring solution in
  five-dimensions}},
  \href{http://dx.doi.org/10.1103/PhysRevLett.88.101101}{\emph{Phys. Rev.
  Lett.} {\bf 88} (2002) 101101},
  [\href{http://arxiv.org/abs/hep-th/0110260}{{\tt hep-th/0110260}}].

\bibitem{Myers:1986un}
R.~C. Myers and M.~J. Perry, \emph{{Black Holes in Higher Dimensional
  Space-Times}},
  \href{http://dx.doi.org/10.1016/0003-4916(86)90186-7}{\emph{Annals Phys.}
  {\bf 172} (1986) 304}.

\bibitem{Elvang:2004rt}
H.~Elvang, R.~Emparan, D.~Mateos and H.~S. Reall, \emph{{A Supersymmetric black
  ring}}, \href{http://dx.doi.org/10.1103/PhysRevLett.93.211302}{\emph{Phys.
  Rev. Lett.} {\bf 93} (2004) 211302},
  [\href{http://arxiv.org/abs/hep-th/0407065}{{\tt hep-th/0407065}}].

\bibitem{Bena:2004wv}
I.~Bena, \emph{{Splitting hairs of the three charge black hole}},
  \href{http://dx.doi.org/10.1103/PhysRevD.70.105018}{\emph{Phys. Rev. D} {\bf
  70} (2004) 105018}, [\href{http://arxiv.org/abs/hep-th/0404073}{{\tt
  hep-th/0404073}}].

\bibitem{Mathur:2018tib}
S.~D. Mathur and D.~Turton, \emph{{The fuzzball nature of two-charge black hole
  microstates}},
  \href{http://dx.doi.org/10.1016/j.nuclphysb.2019.114684}{\emph{Nucl. Phys.}
  {\bf B945} (2019) 114684}, [\href{http://arxiv.org/abs/1811.09647}{{\tt
  1811.09647}}].

\bibitem{Giusto:2004xm}
S.~Giusto and S.~D. Mathur, \emph{{Fuzzball geometries and higher derivative
  corrections for extremal holes}},
  \href{http://dx.doi.org/10.1016/j.nuclphysb.2005.12.012}{\emph{Nucl. Phys.}
  {\bf B738} (2006) 48--75}, [\href{http://arxiv.org/abs/hep-th/0412133}{{\tt
  hep-th/0412133}}].

\bibitem{Papadopoulos:2008rx}
G.~Papadopoulos, \emph{{New half supersymmetric solutions of the heterotic
  string}},
  \href{http://dx.doi.org/10.1088/0264-9381/26/13/135001}{\emph{Class. Quant.
  Grav.} {\bf 26} (2009) 135001}, [\href{http://arxiv.org/abs/0809.1156}{{\tt
  0809.1156}}].

\bibitem{Gran:2005wf}
U.~Gran, P.~Lohrmann and G.~Papadopoulos, \emph{{The Spinorial geometry of
  supersymmetric heterotic string backgrounds}},
  \href{http://dx.doi.org/10.1088/1126-6708/2006/02/063}{\emph{JHEP} {\bf 02}
  (2006) 063}, [\href{http://arxiv.org/abs/hep-th/0510176}{{\tt
  hep-th/0510176}}].

\bibitem{Gran:2007kh}
U.~Gran, G.~Papadopoulos and D.~Roest, \emph{{Supersymmetric heterotic string
  backgrounds}},
  \href{http://dx.doi.org/10.1016/j.physletb.2007.09.024}{\emph{Phys. Lett. B}
  {\bf 656} (2007) 119--126}, [\href{http://arxiv.org/abs/0706.4407}{{\tt
  0706.4407}}].

\bibitem{Fontanella:2019avn}
A.~Fontanella and T.~Ort\'\i{}n, \emph{{On the supersymmetric solutions of the
  Heterotic Superstring effective action}},
  \href{http://dx.doi.org/10.1007/JHEP06(2020)106}{\emph{JHEP} {\bf 06} (2020)
  106}, [\href{http://arxiv.org/abs/1910.08496}{{\tt 1910.08496}}].

\bibitem{Fubini:1985jm}
S.~Fubini and H.~Nicolai, \emph{{The Octonionic Instanton}},
  \href{http://dx.doi.org/10.1016/0370-2693(85)91589-8}{\emph{Phys. Lett. B}
  {\bf 155} (1985) 369--372}.

\bibitem{Gunaydin:1995ku}
M.~Gunaydin and H.~Nicolai, \emph{{Seven-dimensional octonionic Yang-Mills
  instanton and its extension to an heterotic string soliton}},
  \href{http://dx.doi.org/10.1016/0370-2693(95)00375-U}{\emph{Phys. Lett. B}
  {\bf 351} (1995) 169--172}, [\href{http://arxiv.org/abs/hep-th/9502009}{{\tt
  hep-th/9502009}}].

\bibitem{Cano:2017qrq}
P.~A. Cano, P.~Meessen, T.~Ortin and P.~F. Ramirez, \emph{{Non-Abelian black
  holes in string theory}},
  \href{http://dx.doi.org/10.1007/JHEP12(2017)092}{\emph{JHEP} {\bf 12} (2017)
  092}, [\href{http://arxiv.org/abs/1704.01134}{{\tt 1704.01134}}].

\end{thebibliography}\endgroup
\label{biblio}
\end{document}